\documentclass[doublecol]{epl2}
\usepackage{graphicx}
\usepackage{dcolumn}
\usepackage{bm}
\graphicspath{{./}}
\usepackage[total={6.5in,9.75in}, top=0.9in, left=0.9in, includefoot]{geometry}
\usepackage{color}
\usepackage[normalem]{ulem}
\usepackage{xcolor,hyperref}
\usepackage{amsmath}
\usepackage{amssymb}
\usepackage{url}
\makeatletter
\newenvironment{linenomath}{}{}
\newenvironment{linenomath*}{}{}
\makeatother

\newcommand{\orcidicon}{\includegraphics[height=8pt]{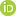}\hspace{1pt}}

\newcommand{\orcidauthorA}{0000-0002-4361-6521}
\newcommand{\orcidauthorB}{0000-0001-7746-2850}
\newcommand{\orcidauthorC}{0009-0008-3270-0702}
\newcommand{\orcidauthorD}{0000-0002-9004-2292}

\newcommand{\orcidA}{\href{https://orcid.org/\orcidauthorA}{\orcidicon}}
\newcommand{\orcidB}{\href{https://orcid.org/\orcidauthorB}{\orcidicon}}
\newcommand{\orcidC}{\href{https://orcid.org/\orcidauthorC}{\orcidicon}}
\newcommand{\orcidD}{\href{https://orcid.org/\orcidauthorD}{\orcidicon}}

\title{Beyond Stokes drift -- Lagrangian transport in evolving gravity waves}
\shorttitle{Beyond Stokes drift -- Lagrangian transport in evolving gravity waves} 

\author{T. Izawa \orcidC{}\inst{1}, G. Foggi Rota \orcidA{}\inst{1}, A. Chiarini \orcidB{}\inst{1,2}, and M. E. Rosti \orcidD{}\inst{1}\thanks{Email for correspondence: marco.rosti@oist.jp}}
\shortauthor{T. Izawa \etal}

\institute{                    
  \inst{1} Complex Fluids and Flows Unit, Okinawa Institute of Science and Technology Graduate University, 1919-1 Tancha, Onna-son, Okinawa 904-0495, Japan.\\
  \inst{2} Dipartimento di Scienze e Tecnologie Aerospaziali, Politecnico di Milano, via La Masa 34, 20156 Milan, Italy.
}

\abstract{
Finite-amplitude gravity waves at the air-water interface induce net fluid and particle transport, known as Stokes drift. While this mechanism is well understood for steady waves, transport under unsteady, evolving conditions remains poorly characterised. Here, we investigate Lagrangian transport in freely decaying waves using high-resolution two-phase simulations and a perturbative analytical model. Wave decay modifies the classical {Lagrangian} drift by introducing both first- and second-order corrections in the wave amplitude expansion, and generates a net vertical transport, governed by the balance between inertia and viscosity. These effects alter particle trajectories and enhance anisotropic mixing, with implications for interpreting field observations and modelling surface transport processes.
}

\begin{document}

\maketitle

\section{Introduction}

More than two-thirds of Earth's surface is covered by oceans, which are essential to sustaining life and regulating the climate system. At the ocean-atmosphere interface, the interaction between winds, currents, and surface waves drives complex exchanges of mass, momentum, and energy, generating motions across a broad range of spatial and temporal scales. On a global scale, oceanic currents shape continental climates \cite{caesar-etal-2018}, while extreme atmospheric events can transport marine aerosols and particulates inland \cite{ryan-etal-2023}. At more local scales, weak winds give rise to Langmuir circulations, near-surface convective cells that accumulate floating material in convergence zones \cite{thorpe-2004}. At the micro-scale, gradients in local viscosity induced by microbial activity and phytoplankton can influence solute transport and the motility of microorganisms \cite{guadayol-etal-2021}.

Surface waves of all wavelengths propagating at the air-water interface \cite{lighthill-2001} play a key role in the transport of sediments \cite{blondeaux-etal-2012} and particulate matter \cite{kaiser-2010,giurgiu-etal-2024}. When waves break, they inject momentum into underlying currents \cite{peregrine-1983,melville-1996,perlin-choi-tian-2013}, enhancing the transport and dispersion of suspended materials. They also facilitate mass, momentum, and energy exchange between the ocean and atmosphere through the formation of bubbles and droplets \cite{digiorgio-pirozzoli-iafrati-2022,deike-2022}.
Even in the absence of breaking, surface gravity waves of finite amplitude induce a net Lagrangian transport—Stokes drift—which causes passive tracers to experience a mean displacement in the direction of wave propagation \cite{stokes-1847,longuett-stoneley-1953,bremer-breivik-2017}. 
Because it emerges from the Lagrangian averaging of wave-induced oscillations, Stokes drift plays a fundamental role in the transport of floating and suspended materials in the upper ocean.
Since its original formulation, the theory has been extended to incorporate more realistic effects, including complex bathymetry \cite{vittori-blondeaux-1996}, nonlinear wave shapes \cite{chang-1969,vandenbroeck-1999,clark-etal-2023,pizzo-etal-2023}, interaction with the Eulerian return flow \cite{vandenbremer-taylor-2016,vandenbremer-etal-2019}, and the presence of tracers with finite size \cite{calvert-etal-2021}, non-spherical geometry \cite{dibenedetto-koseff-ouellette-2019}, inertia \cite{eames-2008,santamaria-etal-2013,deleo-etal-2021,dibenedetto-clark-pujara-2022}, or diffusive behavior \cite{jansons-lythe-1998}.
A less examined but crucial aspect is that wave amplitude itself evolves in time, depending on both the characteristics of the sustaining forcing, such as local wind intensity, and on viscous dissipation, which together influence material transport. In the absence of continuous forcing, wave energy decays over time due to viscosity, leading to a gradual attenuation of wave amplitude \cite{devita-etal-2021,lorenzo-etal-2023}. Although the theory of wave decay was developed from an energetic standpoint by L.D. Landau in the 1930s \cite{landau-lifshitz-2003}, its implications for the net Lagrangian transport induced by unforced, decaying wave fields remain largely unexplored.

\begin{figure}
	\begin{minipage}[t]{0.49\textwidth}
	\raggedright
	\includegraphics[trim={0 3 0 0},clip,width=\textwidth]{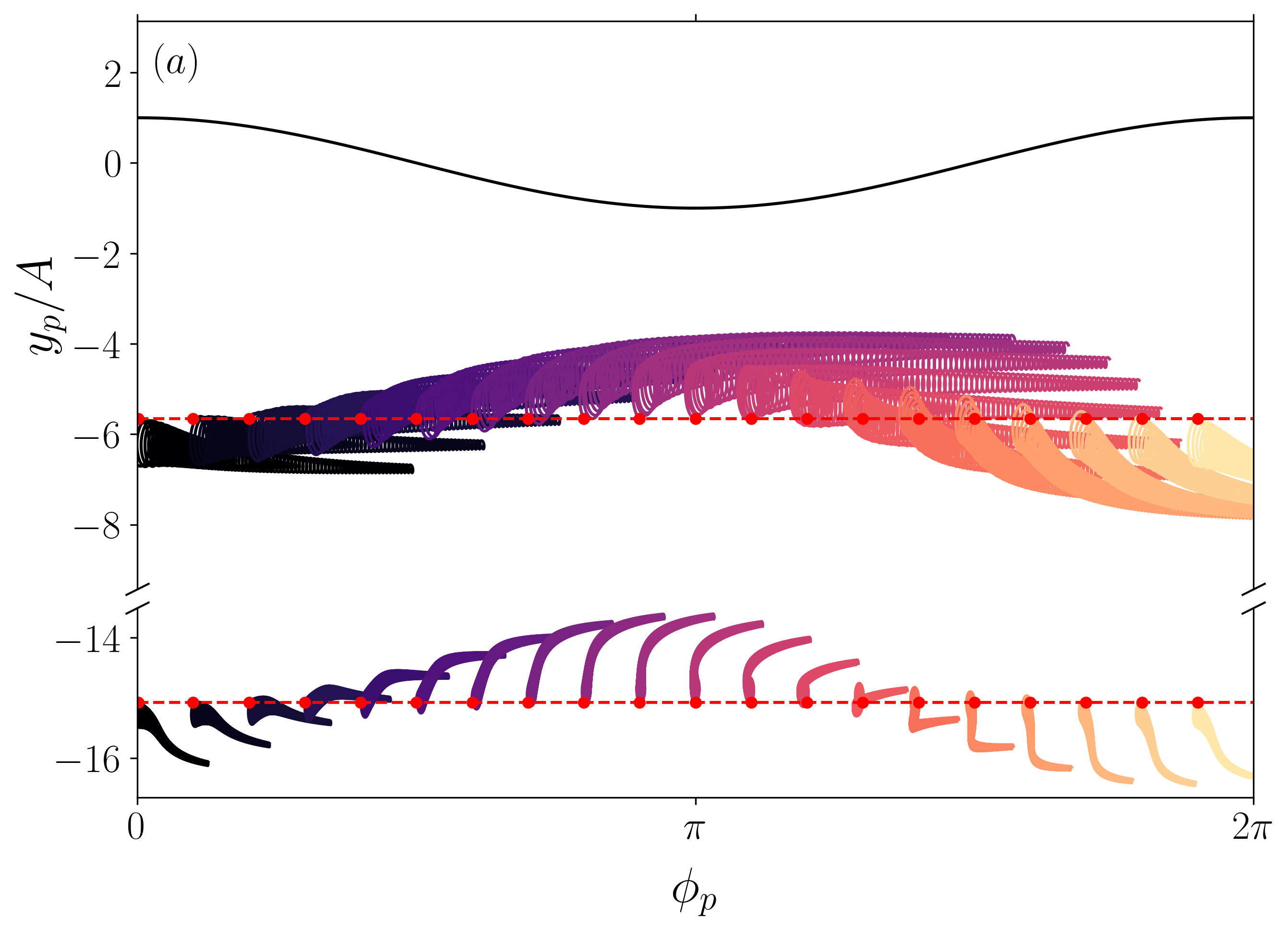}
	\end{minipage}
	\begin{minipage}[t]{0.49\textwidth}
	\raggedright
	\hspace{.05cm}\includegraphics[width=.975\textwidth]{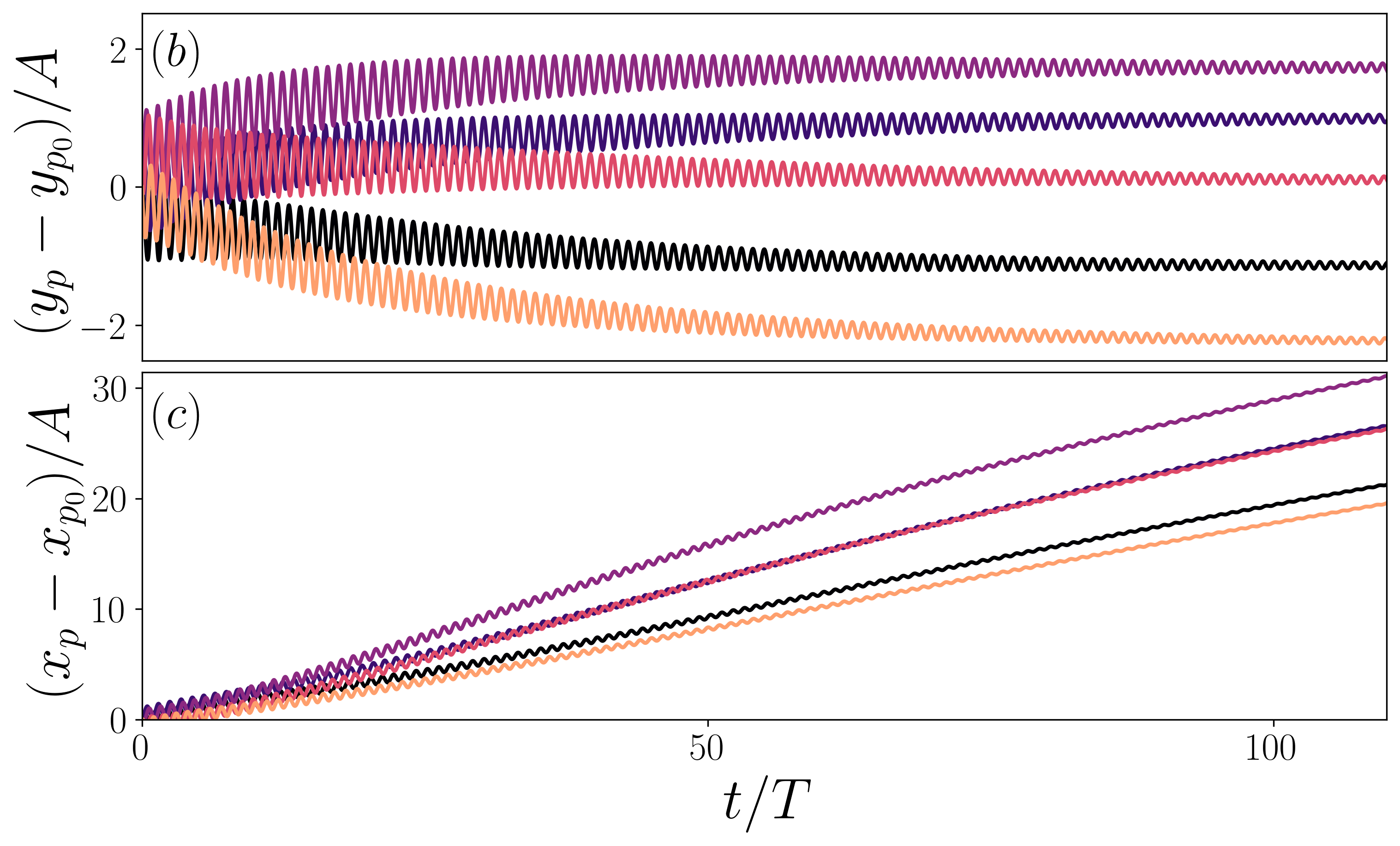}
	\end{minipage}
	\caption{Drift of fluid particles under a decaying wave. (a) Trajectories of particles released at $t=0$ along two longitudinal lines reveal vertical motion beyond classical Stokes drift.  (b) Vertical displacement from the release depth \( y_{p_0}/A = -6 \) depends on the initial phase \(\phi_{p_0}\).  (c) Longitudinal displacement exhibits initial super-linear growth.}
	\label{fig:cover}
\end{figure}

In this Letter, we investigate the net Lagrangian transport induced by freely decaying surface gravity waves in deep water. Using fully nonlinear direct numerical simulations, supported by a perturbative analytical framework, we characterise the transport generated by a monochromatic wave propagating without external forcing and subject to viscous decay. 
Unlike the idealised inviscid case, where Stokes drift yields purely horizontal motion, we thus observe an additional vertical migration of particles (figure  \ref{fig:cover}). Even small vertical displacements cause substantial changes in drift velocity, leading to trajectories that deviate markedly from classical Stokes-drift predictions. In the vanishing-viscosity limit, our analytical model recovers the classical result.

\section{Setup and methods}

We perform direct numerical simulations of the flow induced by a monochromatic surface gravity wave of wavelength $\lambda$ and initial finite amplitude $A$, freely propagating along the air--water interface. Simulations are conducted in a two-dimensional square domain of size $\lambda \times \lambda$, bounded vertically ($y$) and periodic in the horizontal ($ x$) direction. 
{Thus, differently from wave tank setups where boundaries induce recirculation, here nothing prevents the establishment of a global mass flux along the direction of wave propagation, and no return is measured at any depth (more details in the Supplemental Material~\cite{supplementary}).}
The domain is half-filled with water (bottom) and air (top), with the average interface located at $\bar{y}_I = 0$. No-slip and no-penetration conditions are imposed at the {top and bottom walls: far enough from the interface to play a negligible role in the reported statistics (see Supplemental Material~\cite{supplementary})}. This simplified setup clearly differs from realistic conditions, but better allows for the theoretical explanation of the phenomena discussed in the following.

The flow evolves according to the incompressible Navier--Stokes equations:
{
\begin{linenomath}
\begin{align}
\frac{\partial \bm{u}}{\partial t} + \bm{\nabla} \cdot (\bm{u} \bm{u}) &= \frac{1}{\rho_f} \left( - \bm{\nabla} p + \bm{\nabla} \cdot \bm{\tau} + \bm{f}_I \right) + \bm{g}, \\
\bm{\nabla} \cdot \bm{u} &= 0,
\end{align}
\end{linenomath}
}
where $\bm{u}$ is the velocity, $p$ the pressure, $\rho_f$ the local fluid density ($\rho_w $ in water, $\rho_a$ in air), $\bm{\tau}$ the viscous stress tensor, and $\bm{g}$ the gravitational acceleration (pointing downward). The surface tension at the interface is modelled as $\bm{f}_I = \sigma k \delta_I \bm{n} $, where $\sigma$ is the air-water surface tension, $k$ the local curvature, $\delta_I$ a Dirac delta function localised at the interface, and $\bm{n}$ the unit normal.

At $t = 0$, the wave is initialised using the {linear}, inviscid deep-water solution~\cite{lighthill-2001}: $u(x,y,t) = \ U e^{\kappa y} \cos(\kappa x - \omega t)$ and $v(x,y,t) = \ U e^{\kappa y} \sin(\kappa x - \omega t)$, where $ \kappa = 2\pi/\lambda $ is the wavenumber, $ \omega = \sqrt{|\bm{g}| \kappa} $ is the angular frequency, and $ U = \omega A $ is the surface velocity amplitude. This corresponds to an initial interface shape $ y_I(x,t) = A \cos(\kappa x - \omega t) $. The deep-water approximation is valid for $ \kappa h \gtrsim 2 $, where $ h $ is the water depth~\cite{landau-lifshitz-2003}. The wave then evolves freely and decays viscously, maintaining a constant period $T=2 \pi/\omega$ as its amplitude decreases. Corrections to the initial velocity field to account for finite depth \cite{landau-lifshitz-2003} yield negligible differences.

{The wave remains monochromatic with wavelength $\lambda$ throughout the decay, likely as the imposition of periodic boundary conditions only allows for motions of wavelength ${\lambda/n}, \enspace n \in \mathbb{N}$, and the significant water depth adopted casts nonlinear boundary effects far from the evolving interface.
We confirm it by inspecting the interface elevation spectrum $\hat{\eta}(\alpha)$ (figure \ref{fig:decay}, main axes), with a significant contribution from only the first mode $\alpha / \kappa = 1$. 
{Spectral contributions are here scaled at each instant with the value of the first mode $\hat{\eta}(\kappa)$; the evolution of their un-normalised magnitude throughout the decay might be found in the Supplemental Material~\cite{supplementary}. We now}
fit to the interface a sinusoidal profile with fixed $\alpha=\kappa$ at different times, with phase compatible to the initial condition, and measure the evolution of the phase speed $\chi(t)=\varsigma(t)/\kappa$, with $\varsigma(t)$ the measured angular frequency. The measured phase speed $\chi(t)$ deviates less than $1\%$ from the prediction $c=\sqrt{|\bm{g}|/\kappa}$ (figure \ref{fig:decay}, inset), matching the theoretical dispersion within numerical accuracy throughout the decay. Having confirmed that the only active wavenumber throughout the decay is $\kappa$, and that $\chi(t)=c=\sqrt{|\bm{g}|/\kappa}=const.$, necessarily $\varsigma (t)= c \kappa = \sqrt{|\bm{g}| \kappa} =\omega=const.$ and, equivalently, the period stays constant and equal to $T=2\pi / \omega$.
\begin{figure}
\centering
  \includegraphics[width=.4\textwidth]{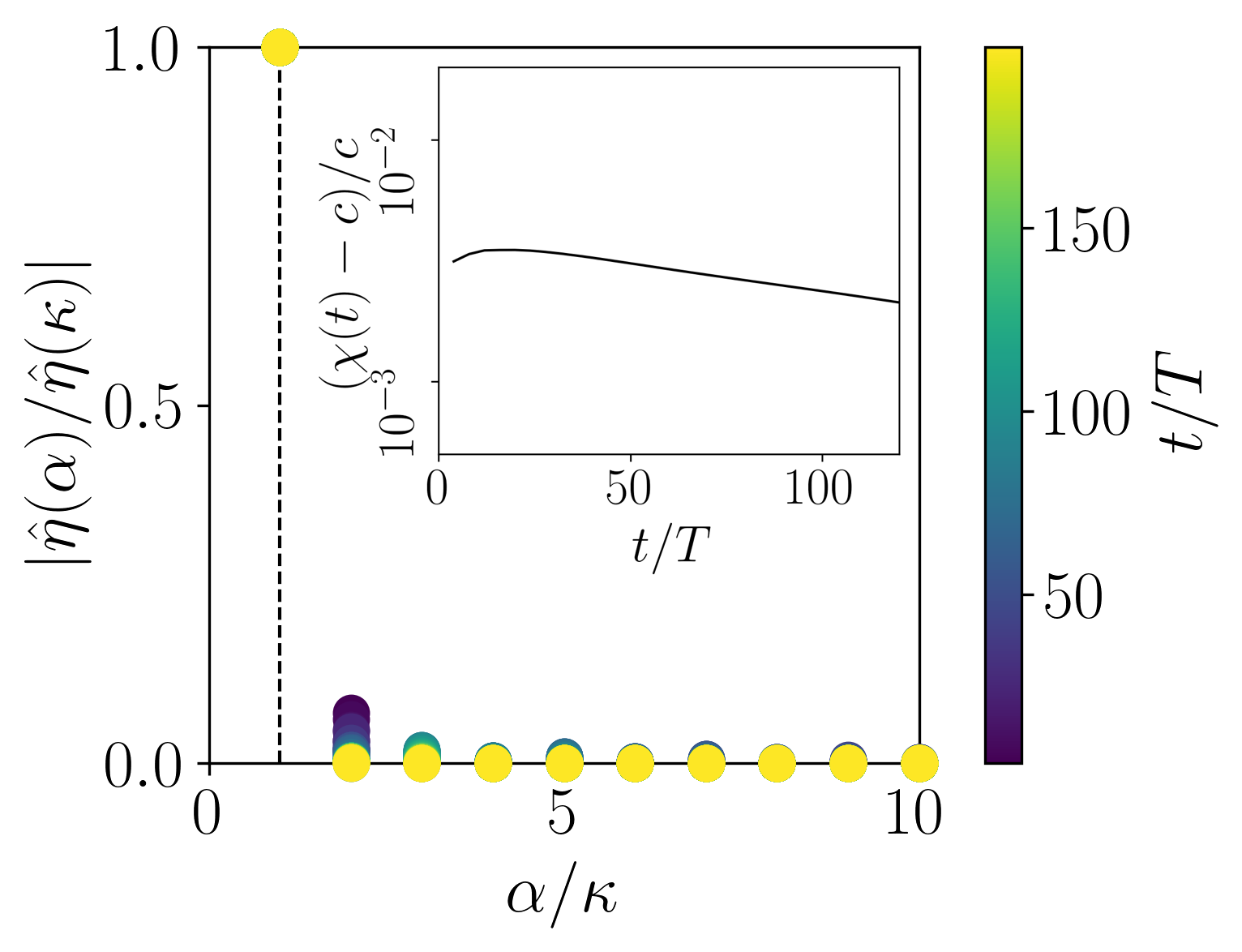}
  \caption{{Wavenumber spectra of the interface elevation $\eta$ at various time instants, from our simulation with $Re=10000$ and $\epsilon=0.1$. The inset reports the deviation of the measured phase speed $\chi(t)$ in that same case from the theoretical prediction, $c=\sqrt{|\bm{g}|/\kappa}$.}}
  \label{fig:decay}
\end{figure}
}

We use realistic air-water parameters: a density ratio $\rho_w/\rho_a = 800$, viscosity ratio $\mu_w/\mu_a = 55$, and surface tension $\sigma$ chosen to yield an { E\"otv\"os number $Eo = \rho_w |\bm{g}| \lambda^2 / \sigma = 12265.5$}: the same physical value found in literature \cite{digiorgio-pirozzoli-iafrati-2022}. Regardless, given the monochromatic and non-breaking nature of the waves considered here, results prove insensitive to arbitrarily high values of {$Eo$} {(likely, dissipative processes at the interface are significantly less relevant than those in the bulk). Yet, within a reasonable range, a finite value of surface tension allows for a more stable integration of the interface dynamics. We thus retain it as a matter of numerical stability.} The initial wave steepness is $\epsilon = \kappa A = 0.1 $, corresponding to non-breaking waves~\cite{digiorgio-pirozzoli-iafrati-2022}. The Reynolds number based on wavelength, $Re = \rho_w \sqrt{|\bm{g}| \lambda^3}/\mu_w$, is first set to $Re=10^4$ and then varied over the range $ Re \in [1000, 12500] $. The water phase is seeded with passive tracer particles that follow the local velocity field. Additional details on the numerical method and computational implementation are provided in the Supplemental Material~\cite{supplementary}.

\section{Results}

The classical Stokes drift associated with inviscid, periodic surface waves predicts purely horizontal (longitudinal) particle motion, with a depth-dependent drift velocity given by $u_d = (U^2 / c) \, e^{2\kappa y}$, where $c = \omega/\kappa$ denotes the wave phase speed. In contrast, under decaying waves, we observe an additional, systematic vertical drift. 
While vertical displacement is a documented effect when particles have inertial properties \cite{santamaria-etal-2013,dibenedetto-clark-pujara-2022}, interact with {an} Eulerian return flow \cite{vandenbremer-taylor-2016,vandenbremer-etal-2019}, {move over an inclined bed \cite{liao-zou-2025}, or in shallow-water waves \cite{grue-kolaas-2017}}, here we discuss a novel and purely kinematic phenomenon arising in freely decaying conditions. The magnitude and direction of the vertical motion depend on the initial phase $\phi_{p_0}$ of particles beneath the wave, as shown in figure \ref{fig:cover}a. This effect is quantified in figure \ref{fig:cover}b, where the vertical displacement saturates after approximately $50$ wave periods, following an initial transient. Simultaneously, the horizontal drift (figure \ref{fig:cover}c) exhibits nonlinear evolution, indicative of a time-dependent drift velocity. Once the particles reach an approximately constant depth, their horizontal displacement becomes nearly linear in time, recovering a classical Stokes-like drift despite the ongoing decay of the wave amplitude.
{
A physical interpretation of the vertical drift is as follows. For infinitesimal-amplitude, non-decaying waves in deep water, fluid particles follow closed circular orbits~\cite{lighthill-2001}. At finite, steady amplitude, these orbits become trochoidal, leading to a net horizontal displacement as a nonlinear effect.
Finite viscosity alters the particle response. In this case, the wave amplitude decays over time, causing the particle orbits to progressively shrink. As a result, a particle moving upward (or downward) and downstream (or upstream) will return along a shorter path in the opposite direction, due to the reduced wave amplitude. Since the wave period remains unchanged, this asymmetry accumulates over successive cycles, producing a net drift in both vertical and horizontal directions, even at infinitesimal amplitude. The direction of these drifts is determined by the particle’s initial phase beneath the wave. As shown below, viscosity modifies the Lagrangian drift by introducing a linear contribution in $A$ and altering the classical Stokes drift.
}

To examine the effect of wave decay on the longitudinal and vertical displacements of the particles, we develop an analytical model based on a perturbative expansion in wave steepness. We focus on the deep-water limit, $\kappa h \rightarrow \infty$, and assume, following Landau~\cite{landau-lifshitz-2003}, that $\nu_w \ll \omega/\kappa^2$. Under this assumption, vorticity is confined to a thin surface layer near the interface, allowing the bulk flow to be treated as irrotational. The wave amplitude decays exponentially as $A e^{-\gamma t}$, with damping coefficient $\gamma = 2 \nu_w \kappa^2$. Enforcing the small-steepness condition {$\epsilon = 2\pi A / \lambda \ll 1$}, equivalent to a small Froude number $\mathrm{Fr} = U/c$, ensures that the velocity field retains its functional form and decays self-similarly.
We analyse the motion of passive fluid particles governed by $\text{d}\bm{x}_p/\text{d}t = \bm{u}_p(t) = \bm{u}(\bm{x}_p(t), t)  $ with initial position $ \bm{x}_p(0) = \bm{x}_{p_0}$. By performing a Taylor expansion of $\bm{u}(\bm{x}_p(t), t)$ around $\bm{x}_{p_0} $ and expanding the particle position in powers of $\epsilon$, we derive a hierarchy of equations solvable order by order (see Supplemental Material~\cite{supplementary}).
Truncating the expansion at $\mathcal{O}(\epsilon^2)$, we find the net horizontal and vertical local tracer velocities
\begin{align}
u_p(t) &= U e^{-\gamma t} e^{\kappa y_{p_0}} \cos(\kappa x_{p_0} - \omega t) \notag \\
&\quad + \frac{U^2 \kappa}{\omega} \frac{e^{2\kappa y_{p_0}}}{(\gamma/\omega)^2 + 1} 
\Big[ e^{-2\gamma t} \notag \\
&\quad + { e^{-\gamma t} \left( \frac{\gamma}{\omega} \sin(\omega t) - \cos(\omega t) \right) } \Big]
, \label{eq:exp-u} \\
v_p(t) &= U e^{-\gamma t} e^{\kappa y_{p_0}} \sin(\kappa x_{p_0} - \omega t) \notag \\
&\quad + \frac{U^2 \kappa}{\omega} \frac{e^{2\kappa y_{p_0}}}{(\gamma/\omega)^2 + 1} 
\Big[ - \frac{\gamma}{\omega} e^{-2\gamma t} \notag \\
&\quad + { e^{-\gamma t} \left( \frac{\gamma}{\omega} \cos(\omega t) {+} \sin(\omega t) \right) } \Big]. \label{eq:exp-v}
\end{align}
In both equations, the first and second terms on the right-hand side correspond to contributions at orders $\epsilon$ and $\epsilon^2$, respectively.
In the inviscid limit ($\gamma = 0$), averaging over one wave period $T$ eliminates the oscillatory term in equation~\eqref{eq:exp-u} and all terms in equation~\eqref{eq:exp-v}, leaving only the classical horizontal Stokes drift velocity, $(u_d, v_d) = (1/T) \int_{t_0}^{t_0+T} (u_p, v_p)\mathrm{d}t = \left( (U^2/c) e^{2\kappa y_{p_0}}, 0\right)$. As expected, the Stokes drift is a second-order effect in wave amplitude. 
{
When wave decay is present ($\gamma > 0$), both first- and second-order contributions survive after averaging and contribute to the Lagrangian drift velocities (see Supplemental Material~\cite{supplementary}). Consequently, decaying waves induce time-dependent drift in both horizontal and vertical directions, reflecting the changing wave amplitude. 
{The first-order terms, which are absent in the inviscid case ($\gamma=0$), originate from the leading-order Eulerian velocity field and emerge upon temporal integration due to viscous damping.}
Equation \eqref{eq:exp-u} shows that viscosity influences both contributions to the horizontal drift. Specifically, the first-order term can result in either upstream or downstream motion, depending on the particle's initial phase. The second-order term corresponds to the viscosity-modified Stokes drift, which, in the limit $\gamma \to 0$, reduces to a purely downstream drift for all phases.
Equation \eqref{eq:exp-v} further reveals the emergence of a vertical drift, entirely absent in the inviscid limit, with both first- and second-order components contributing. These can induce either upward or downward motion depending on the phase, while the overall magnitude decreases with increasing Reynolds number (i.e., as $\gamma \to 0$).
Overall, the model shows that both horizontal and vertical drift velocities depend sensitively on the particle's initial phase, in agreement with the numerical results in figure \ref{fig:cover}. Full derivation, validation, and comparison with direct numerical simulations are provided in the Supplemental Material~\cite{supplementary}.
}

\begin{figure}
	\centering
	\includegraphics[width=.49\textwidth]{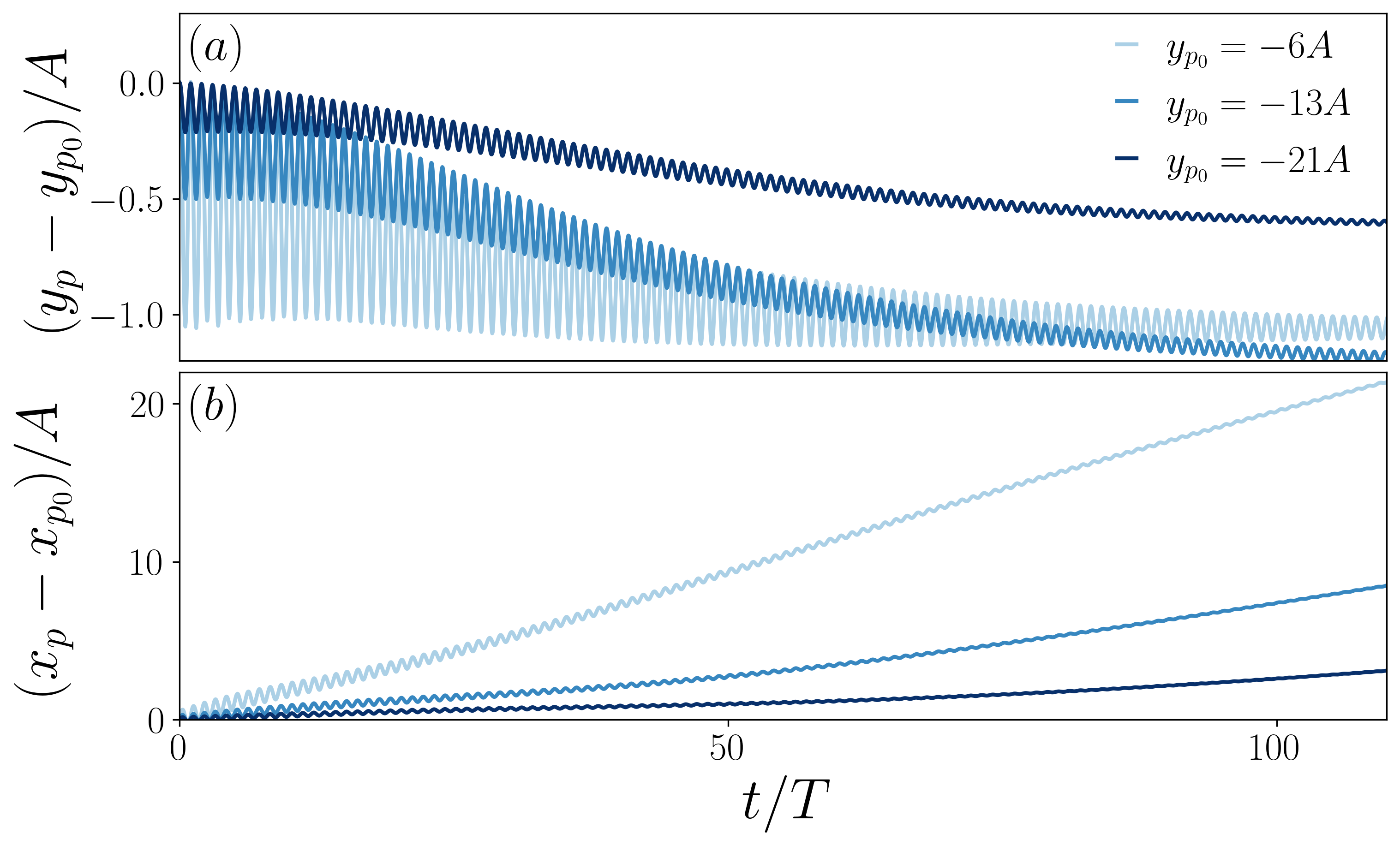}
	\caption{Vertical $(a)$ and longitudinal $(b)$ displacement of fluid particles released under the wave at the same phase $\phi_{p_0}=0$, but different depths.}
	\label{fig:depthEffect}
\end{figure}

This seemingly simple mechanism has profound implications for particle dynamics, due to its nonlinear interaction with the depth-dependent velocity field beneath the wave. In particular, it strongly influences the mixing properties of the flow, which we now examine. Based on our direct numerical simulations, we consider fluid particles initially released along a vertical line beneath the air-water interface (e.g., arbitrarily choosing an initial phase $\phi_{p_0}=0$). Since all particles share the same initial phase, they drift coherently in the same vertical direction (i.e., downward, for the choice of $\phi_{p_0}=0$ shown in figure \ref{fig:depthEffect}a). However, due to the exponential decay of velocity with depth, particles closer to the interface initially move faster, as seen in figure \ref{fig:depthEffect}b.
This spatial decay interacts nontrivially with the temporal decay of the wave. A particle near the surface descends more rapidly at first, but quickly enters a region of weaker flow. Conversely, a particle released slightly deeper experiences slower vertical initial motion but sustains it longer, ultimately undergoing a larger vertical net displacement. This counterintuitive behaviour is illustrated by the overlapping trajectories in figure \ref{fig:depthEffect}a and reveals a novel mixing mechanism, which we quantify using pair dispersion. 

\begin{figure}
  \centering
  \includegraphics[width=.49\textwidth]{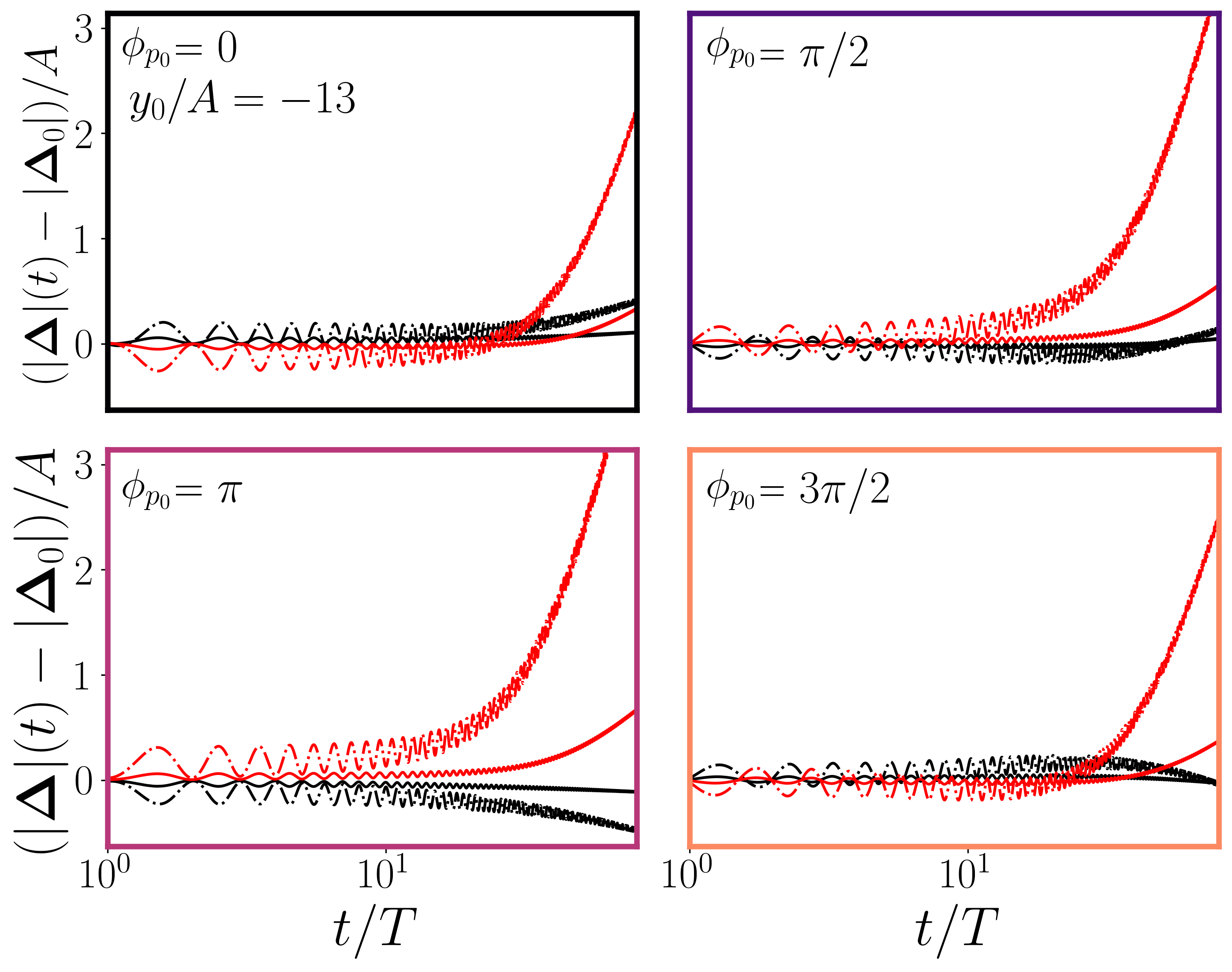}
  \caption{Pair dispersion for $Re = 1000$ and $\epsilon = 0.1$. Black and red curves correspond to particle pairs initially separated in the longitudinal ($x$) and vertical ($y$) directions, respectively. The solid and dash-dotted lines represent initial separations $\Delta_{x_p/y_p,0} = \lambda/64$ and $\lambda/16$, respectively. The strongest dispersion is observed for vertically separated particles, indicating enhanced mixing in that direction. However, a non-negligible dispersion is also induced by longitudinal separations due to the decay of the wave.}
  \label{fig:pair-disp}
\end{figure}

Let us consider particle pairs initially separated either in the longitudinal ($x$) or vertical ($y$) direction, with initial separations {$\Delta_{x,0} = \lambda/64$ or $\Delta_{y,0} = \lambda/16$}. The evolution of the distance between the two particles is quantified as $\Delta(t) = |(\Delta_x(t), \Delta_y(t))| =  |\bm{x}_{p,2}(t) - \bm{x}_{p,1}(t)|$. Figure \ref{fig:pair-disp} shows results for particle pairs whose midpoint is initially located at $y_{p_0} = -13A$, and at four different horizontal positions corresponding to $\phi_{p_0}=\kappa x_{p_0} = 0$, $\pi/2$, $\pi$, and $3\pi/2$. Pairs initially aligned in the vertical direction exhibit significantly stronger dispersion due to the differential drift experienced by the two particles. In contrast, longitudinally aligned pairs remain much closer over time, indicating weaker mixing. The mixing behaviour also depends on the initial phase of the wave at the particle location. For vertically separated pairs, the phase affects the magnitude of the separation but does not change the qualitative trend. However, for longitudinally separated pairs, the initial phase can lead to either convergence or divergence of the particle trajectories as time evolves.
Overall, vertically aligned particles are rapidly separated due to differences in net drift, whereas longitudinally aligned particles experience similar drift and remain closer together. Interestingly, even for the latter, pair dispersion is finite and varies with the phase, reflecting the trajectory patterns observed in figure \ref{fig:cover}a.

\begin{figure}
  \centering
  \includegraphics[width=.49\textwidth]{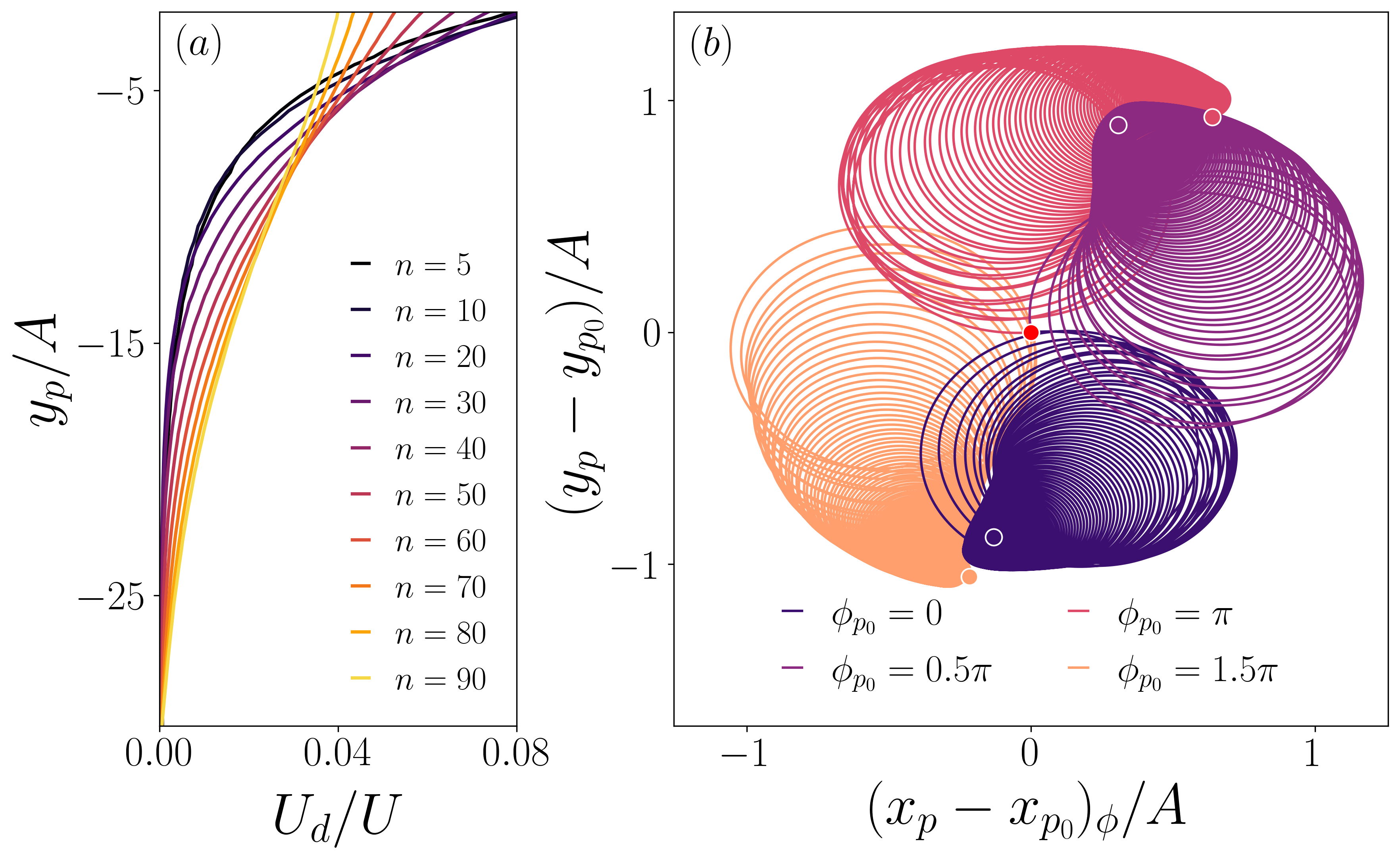}
  \caption{(a) Time-evolving profile of the phase- and period-averaged particle velocity, defined as {$U_d(y_p,n) = \frac{1}{\lambda T} \int_0^\lambda \int_{t_0+nT}^{t_0+(n+1)T} u_p(t;x_{p_n},y_{p_n}=y_p) \text{d}x_{p_n} \text{d} t$}, after binning the vertical coordinate. (b) Fluid particle trajectories with $U_d(y_{p_0},n)$ removed from the longitudinal velocity at each $t$ and $y$, isolating the phase-dependent motion. Particles are released at depth $y_{p_0}/A = -6$.}
  \label{fig:drift}
\end{figure}


We now isolate the influence of the initial phase on fluid-particle trajectories. To this end, we uniformly seed the water phase with passive tracers and compute the phase- and period-averaged particle velocity, {$U_d(y_p,n) = \frac{1}{\lambda T} \int_0^\lambda \int_{t_0+nT}^{t_0+(n+1)T} u_p(t;x_{p_n},y_{p_n}=y_p) \text{d}x_{p_n} \text{d} t$, where $\bm{x}_{p_n}$ denotes the tracer position at time $t = t_0+ nT$}, and show it in figure \ref{fig:drift}(a) after binning over the vertical coordinate.
In the inviscid limit ($\gamma = 0$), $U_d$ reduces to the classical Stokes drift $u_d$. 
As the wave decays, the profile $U_d(y_p,n)$ evolves in time: it decreases near the interface, increases at intermediate depths, and eventually vanishes as the wave is fully damped.
We then recompute the particle trajectories after subtracting the $ U_d(y_p, n)$ value from the longitudinal velocity at each depth and time $t \in [nT, (n+1)T)$, thereby isolating the phase-dependent motion. 
For a steady wave of finite amplitude, the corrected trajectories form closed orbits. 
In the decaying case, the corrected trajectories---denoted by the subscript \(\phi\)---are shown in figure \ref{fig:drift}(b) for four initial phases, with tracers released at $y_{p_0} / A = -6$. Each trajectory starts at the central red marker and ends at a colour-coded point indicating its initial phase.
As expected, wave decay causes the trajectories to shrink over time. Notably, the centres of rotation also drift both horizontally and vertically, with the net offset dependent on the initial phase. After the wave has fully decayed, particles remain displaced in both directions. While the longitudinal offset is small compared to the total drift, the vertical displacement leads particles to sample different values of $U_d$, ultimately resulting in significant deviations from the steady-wave case.

\begin{figure}
	\centering
	\includegraphics[width=.49\textwidth]{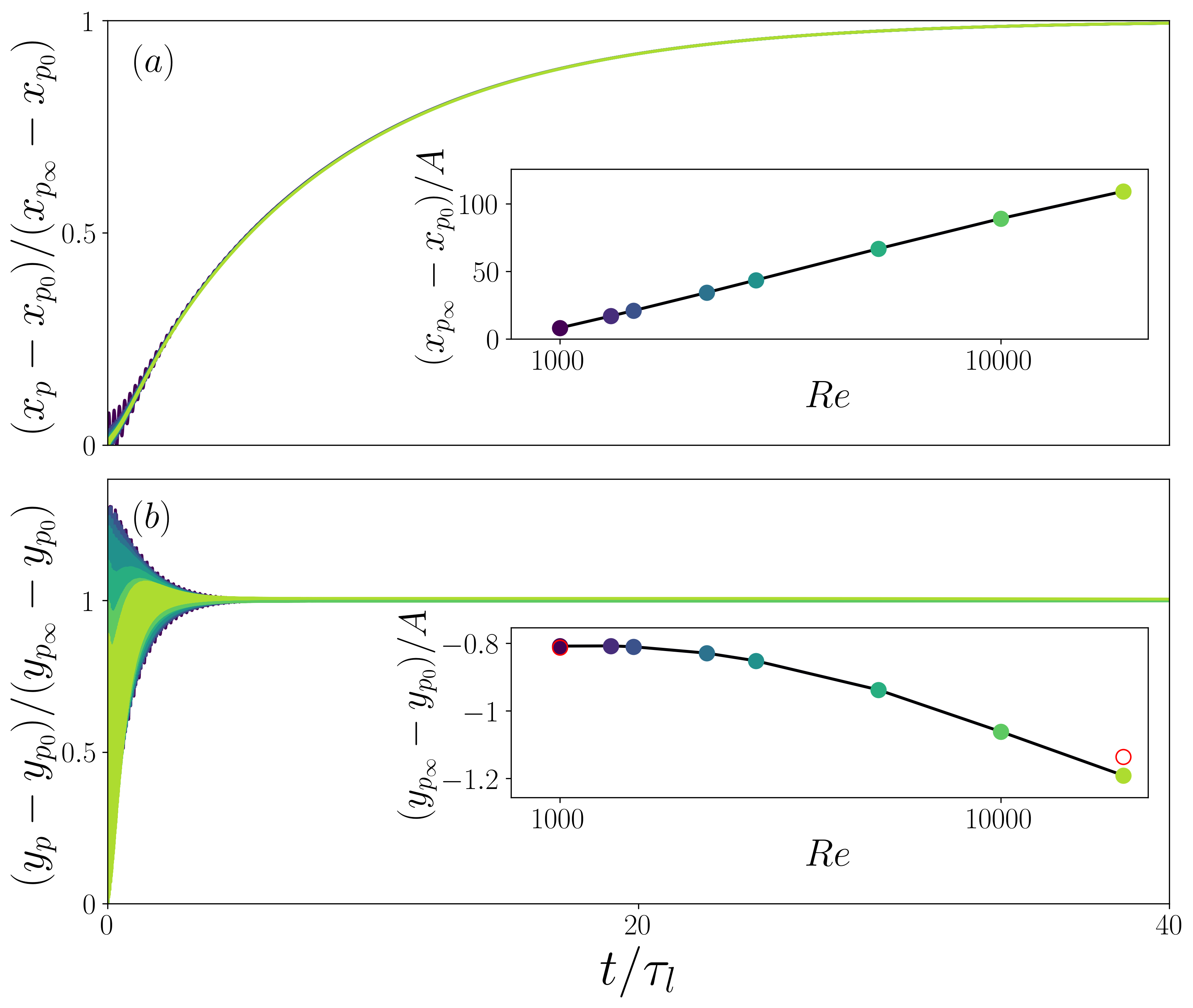}
	\caption{Reynolds number dependence of particle displacement in the longitudinal (a) and vertical (b) directions for a particle released at phase $\phi_{p_0} = 0$ and depth $y_{p_0}/A = -6$. Insets show the final displacements after wave decay across different Reynolds numbers, serving as colour legends for the main curves. The qualitative behaviour remains consistent for varying value of $\phi_{p_0}$ and $y_{p_0}$. In the inset of panel (b), we also report as red circles the results attained for two selected cases doubling the spatial resolution, noting minimal deviation from the main study.}
	\label{fig:reTrend}
\end{figure}
The mechanism identified in figure \ref{fig:drift}b, and analytically described by equations \eqref{eq:exp-u}-\eqref{eq:exp-v}, underpins the anomalous particle dynamics observed in decaying waves. To assess its physical relevance, we examine its dependence on the Reynolds number by tracking the long-time displacement ($ t \rightarrow \infty $) of a fluid particle released at phase $\phi_{p_0} = 0$ and depth $y_{p_0} / A = -6$ {at $t=t_0$}, across increasing values of $Re$.
Remarkably, when displacements are normalised by their asymptotic value and time is rescaled by the viscous decay timescale $\tau_l = 1/\gamma$, all trajectories collapse onto a universal curve (main panels of figure \ref{fig:reTrend}). The inset of figure \ref{fig:reTrend}a shows that the asymptotic longitudinal displacement $(x_{p_\infty}-x_{p_0})$ scales linearly with $Re$, i.e., $ \lim_{t \rightarrow \infty} (x_{p}(t)-x_{p_0}) / A \sim Re^1$. In contrast, the inset of figure \ref{fig:reTrend}b suggests that the (modulus of the) vertical displacement $(y_{p_\infty}-y_{p_0})$ grows with $Re$ only beyond a threshold $Re \gtrsim 5000$, potentially also following a linear trend.
The longitudinal scaling is qualitatively well captured by our analytical model. Integrating equation \eqref{eq:exp-u} in time, and taking the limits $t \rightarrow \infty$ followed by $\gamma \rightarrow 0$ ($Re \rightarrow \infty$), yields:
\begin{linenomath}
\begin{equation}
\begin{aligned}
  x_{p_\infty}-x_{p_0} & = \frac{U}{\omega} e^{\kappa y_{p_0}} \sin(\kappa x_{p_0}{-\omega t_0}) \\
                       &+ \frac{U^2 \kappa}{\omega} e^{2 \kappa y_{p_0}}
   \left[ \frac{1}{2 \gamma} + {\frac{\sin(\omega t_0)}{\omega}} \right] \sim Re.
\end{aligned}
\end{equation}
\end{linenomath}
By contrast, the model predicts that the vertical displacement (from equation \eqref{eq:exp-v}) approaches a finite value determined solely by the particle's initial position, when the limits are taken in the same order (see the Supplemental Material~\cite{supplementary}). Of course, reversing the order---first $Re \rightarrow \infty$, then $t \rightarrow \infty$---recovers the inviscid limit, in which $(y_{p_\infty}-y_{p_0}) \rightarrow 0$. The discrepancy observed at high $Re$ between the model and simulations, robust to grid refinement, likely arises from neglected nonlinearities, finite-amplitude effects beyond the reach of the perturbative expansion, or inherent limitations of the Landau framework. Further discussion is provided in the Supplemental Material~\cite{supplementary}.

\section{Discussion}

We conjecture that the drift mechanism identified in this Letter may have escaped detection in previous studies because most field and wave-tank measurements focus on tracers under passing wave packets. Such configurations do not capture the cumulative effect of wave decay highlighted here, which instead emerges when tracers remain exposed to the same wave field as it decays in time, as occurs in our simulations owing to the use of periodic boundary conditions.
To isolate this mechanism experimentally, we propose wave-tank studies in which tracers are subjected to incident waves with gradually decreasing amplitude. The drift reported here may be particularly relevant in older sea states, where waves are no longer actively forced, and should therefore be considered when interpreting oceanic observations and modelling transport processes in the near-surface layer.

{Importantly, the vertical displacement identified in this study originates solely from the temporal decay of the wave field and is thus distinct from other vertical drift mechanisms discussed in the literature, such as inertia-driven tracer effects \cite{santamaria-etal-2013} or wave-induced Eulerian return flows \cite{vandenbremer-etal-2019}.}


Plastic residues from industrial activity and mismanaged waste enter the ocean in vast quantities every day, contaminating the global water cycle. 
Contrary to popular depictions of plastic pollution as floating debris fields, the reality is dominated by a diffuse suspension of microscopic particles~\cite{kaiser-2010}, which pose severe risks to marine organisms and ecosystems \cite{cui-etal-2023}. The long-term fate of these microplastics remains poorly understood: they may fragment into nanoplastics, settle into sediments, or be carried to coastal regions.
The kinematic drifting mechanism analysed in this Letter 
might offer new insight into these open questions. 

%
%

\acknowledgments
The research was supported by the Okinawa Institute of Science and Technology Graduate University (OIST) with subsidy funding to M.E.R. from the Cabinet Office, Government of Japan. M.E.R. also acknowledges funding from the Japan Society for the Promotion of Science (JSPS), grants 24K17210 and 24K00810. The authors acknowledge the computer time provided by the Scientific Computing \& Data Analysis section of the Core Facilities at OIST, and by HPCI, under the Research Project grants hp230018, hp240006, and hp250035.
\\
\\
All data needed to evaluate the conclusions are present in the manuscript and/or the Supplemental Material~\cite{supplementary}.
Data required to reproduce the figures are available on the website of the Complex Fluids and Flows Unit at OIST (\url{https://www.oist.jp/research/research-units/cffu/publications/publication-data}). Numerical simulations were performed using a standard direct numerical simulation solver for the Navier–Stokes equations.  
Details on the code implementation and validation are available at \url{www.oist.jp/research/research-units/cffu/fujin} {and published \cite{rosti-2026}}.

%
%

\bibliographystyle{eplbib}

\end{document}


\title{SUPPLEMENTAL MATERIAL\\[8pt]
Beyond Stokes drift -- Lagrangian transport in evolving gravity waves}

\author{Tatsuo Izawa \orcidC{}$^1$}
\author{Giulio Foggi Rota \orcidA{}$^1$}
\author{Alessandro Chiarini \orcidB{}$^{1,2}$}
\author{Marco Edoardo Rosti \orcidD{}$^1$}
\email[E-mail for correspondence: ]{marco.rosti@oist.jp}
\affiliation{$^1$ Complex Fluids and Flows Unit, Okinawa Institute of Science and Technology Graduate University, 1919-1 Tancha, Onna-son, Okinawa 904-0495, Japan.\\ $^2$ Dipartimento di Scienze e Tecnologie Aerospaziali, Politecnico di Milano, via La Masa 34, 20156 Milan, Italy.}

\maketitle

\onecolumngrid

{
\section{The numerical simulations}
}
\label{sec:methods}

{
\subsection{Numerical methods}
}

We perform direct numerical simulations in a two-dimensional domain of length $\lambda$ in the longitudinal ($x$) direction, aligned with the wave propagation, and height $\lambda$ in the vertical ($y$) direction. The domain is half-filled with water and subject to gravity $\bm{g} = (0, -g)$ acting along the negative $y$ axis.
The {top and bottom boundaries are} modelled as a solid wall with no-slip and no-penetration conditions. 
Periodic boundary conditions are imposed in the longitudinal direction. The origin of the Cartesian coordinate system is placed at the undisturbed interface between the two fluids when the system is at rest.

The flow is governed by the incompressible Navier--Stokes equations:
\begin{align}
\frac{\partial \bm{u}}{\partial t} + \bm{\nabla} \cdot (\bm{u} \bm{u}) &= \frac{1}{\rho_f} \left( -\bm{\nabla} p + \bm{\nabla} \cdot \bm{\tau} { + \bm{f}_I } \right) + \bm{g}, \enspace \bm{\nabla} \cdot \bm{u} = 0, 
\end{align}
where $\bm{u}$ and $p$ denote the velocity and pressure fields, respectively, and $\rho_f$ is the local fluid density—$\rho_w$ in water and $\rho_a$ in air. The stress tensor is denoted by $\bm{\tau}$. The interfacial force $\bm{f}_I$ accounts for surface tension effects at the interface between the two fluids, where the velocity field is continuous but the stress tensor exhibits a jump. This surface tension force is modeled as a volumetric force:
\begin{equation}
\bm{f}_I = \sigma k \delta_I \bm{n},
\end{equation}
where $\sigma$ is the surface tension coefficient, $k$ is the local curvature of the interface, $\bm{n}$ is the unit normal vector at the interface, and $\delta_I$ is a Dirac delta function nonzero only at the interface.

The motion of Lagrangian tracers is governed by:
\begin{equation}
\frac{\mathrm{d} \bm{x}_p}{\mathrm{d} t} = \bm{u}(\bm{x}_p(t), t),
\label{eq:partDyn}
\end{equation}
where $\bm{x}_p(t)$ denotes the tracer position at time $t$, and $\bm{u}(\bm{x}_p(t), t)$ is the interpolated fluid velocity at that position and time.
{
By doing so, {particles are ``blind'' to the fluid phase in which they lie}, and might drift away from the interface even if initially released on it.
Our numerics is thus unable to track particles constrained to remain on the surface (indeed, here we always considered particles at a finite submergence depth). The problem of floating particles {bound} to the interface is nevertheless relevant and physically backed, e.g., by applications to floating debris in the oceans: future works could address it introducing a penalty term in equation~\ref{eq:partDyn} to constrain particle motion at the surface.
}

Simulations are initialised using the inviscid, {linear}, deep-water wave solution:
\begin{equation}
u(x, y, t) = U e^{\kappa y} \cos(\kappa x - \omega t), \qquad v(x, y, t) = U e^{\kappa y} \sin(\kappa x - \omega t),
\end{equation}
where $\kappa = 2\pi/\lambda$ is the wavenumber, $\omega = \sqrt{g \kappa}$ is the angular frequency, and $U = \omega A$ is the surface velocity amplitude related to the initial wave amplitude $A$. This corresponds to an initial interface shape:
\begin{equation}
y_I = A \cos(\kappa x - \omega t).
\end{equation}
For $t > 0$, the wave is allowed to freely evolve and decay due to viscous dissipation, with a constant period $T = 2\pi/\omega$.

We employ representative air-water parameters. The density and viscosity ratios are set to $\rho_w/\rho_a = 800$ and $\mu_w/\mu_a = 55$, respectively. The surface tension is chosen to yield a realistic E\"otv\"os number $Eo = \rho_w g \lambda^2/\sigma = 12265.5$. The initial wave steepness is $\epsilon \equiv \kappa A = 0.1$, and the Reynolds number is defined as
\begin{equation}
Re = \frac{\rho_w \sqrt{g \lambda^3}}{\mu_w},
\end{equation}
which is varied in the range $1000 \leq Re \leq 12500$.

The governing equations are solved using the in-house solver \textit{Fujin} (\url{https://www.oist.jp/research/research-units/cffu/fujin}), which employs a fractional-step method. The Poisson equation, used to enforce incompressibility, is solved with a fast Poisson solver. Spatial derivatives are discretised using a second-order central finite-difference scheme, and the time integration is performed using a second-order Adams--Bashforth method. Flow variables are stored on a staggered Cartesian grid with $256$ grid points uniformly distributed along both the $x$ and $y$ directions. For selected Reynolds numbers, we performed additional simulations at doubled grid resolution for validation purposes. The Lagrangian tracer equation is advanced in time using the same second-order Adams–Bashforth scheme, {with two sub-iterations for each step of the flow solver}. At each step, the velocity at tracer positions is computed via second-order linear interpolation from the Eulerian grid.

{
The volume-of-fluid method \cite{hirt-nichols-1981} adopted to tackle numerically the multiphase system at hand allows us to solve only one set of equations over the whole domain. This is achieved by introducing a velocity vector field $\mathbf{u}$ valid everywhere, found applying the volume averaging procedure \cite{quintard-whitaker-1994a}
\begin{equation}
	\mathbf{u}=(1-\phi)\mathbf{u_a}+\phi \mathbf{u_w},
\end{equation}
where $\phi$ is the volume-of-fluid function, which is $1$ in water and $0$ in air. In particular, the isoline at $\phi=1/2$ represents the air-water interface. 
The stress tensor $\tau$ is also written in mixture form, similarly to the velocity field above.
To close the full system of Eqns.~(1,2) in the main text, one additional transport equation is needed for the VOF function $\phi$,
\begin{equation}
	\displaystyle \frac{\partial \phi}{\partial t}+\nabla \cdot (\mathbf{u}\mathscr{H})=\phi \nabla \cdot \mathbf{u},
\end{equation}
where $\mathscr{H}$ is a colour function related to the numerical VOF function $\phi$ through a cell-average procedure.  The adopted VOF formulation follows the approach introduced in \cite{li-etal-2012}. 
In particular, by applying this method we impose no constraint on the velocity at the interface nor on the stress, which jumps between air and water of an amount equal to the surface tension contribution.
}
The numerical method has been previously validated for decaying wave simulations through comparison with literature data on wave energy decay, see \citet{foggirota-chiarini-rosti-2025}. \\

\begin{figure}
\centering
  \includegraphics[width=.3\textwidth]{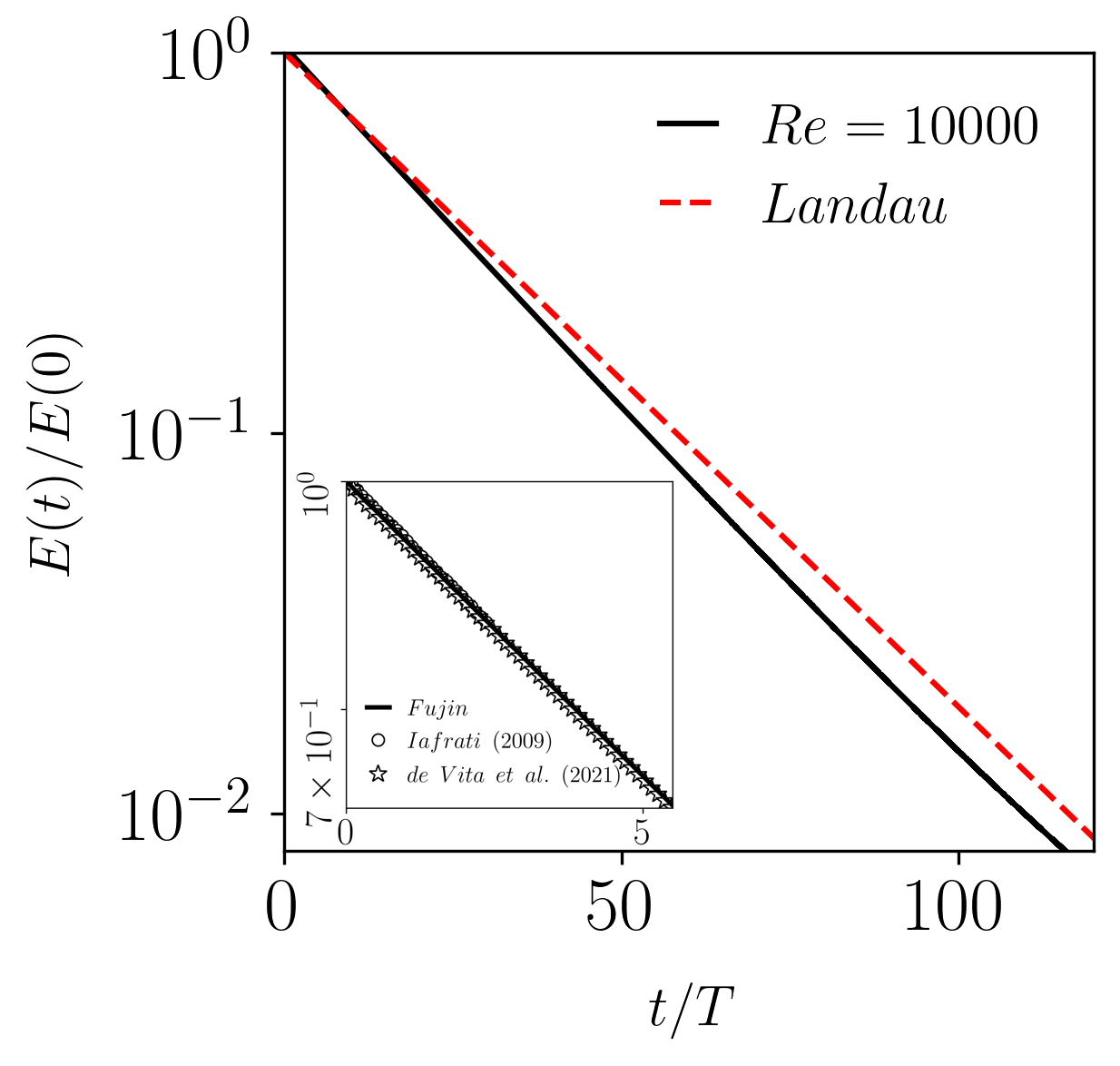}
  \hspace{.5 cm}
  \raisebox{.93 cm}{\includegraphics[width=.215\textwidth]{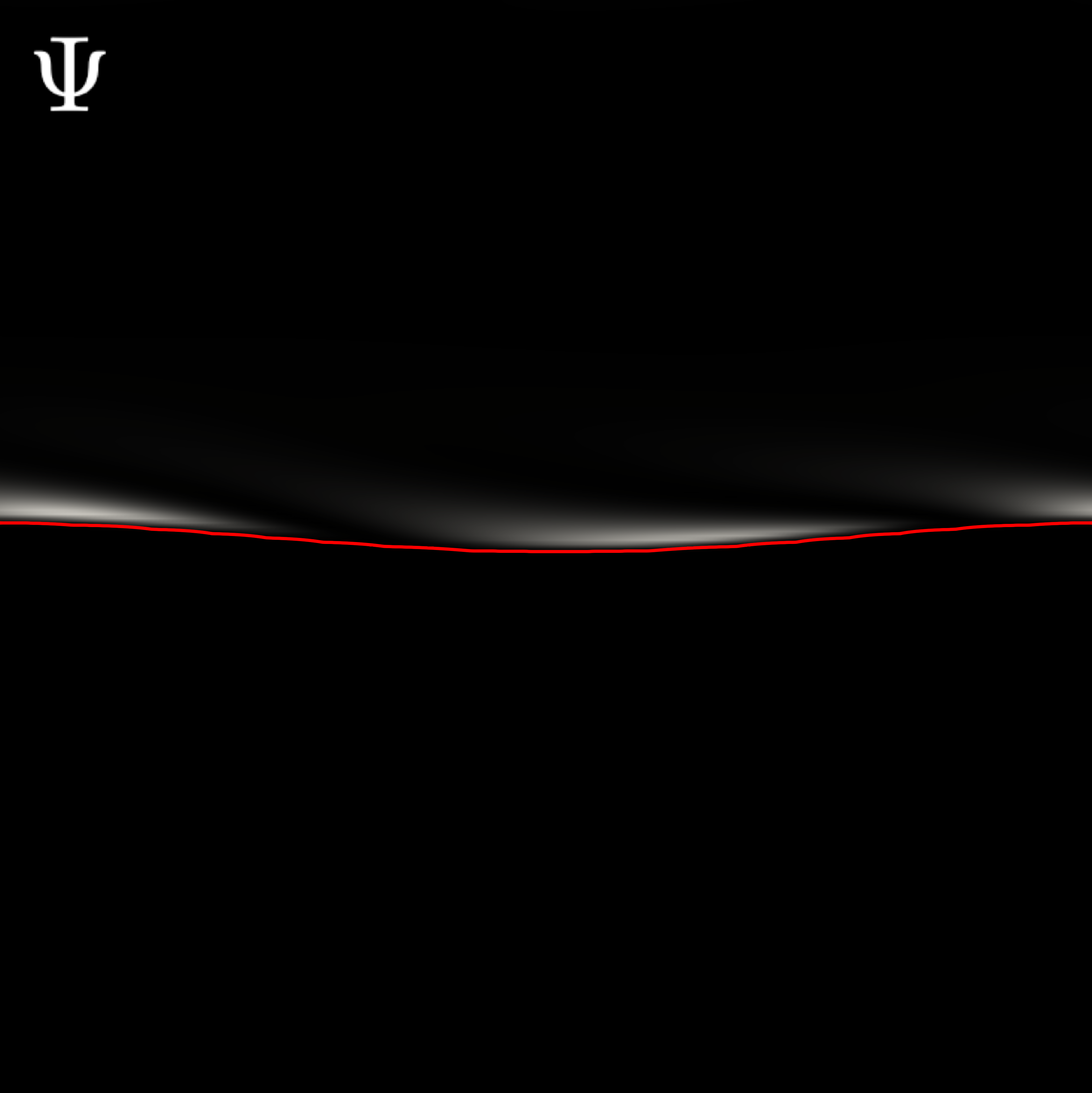}}
  \hspace{.1 cm}
  \raisebox{.2 cm}{\includegraphics[width=.34\textwidth]{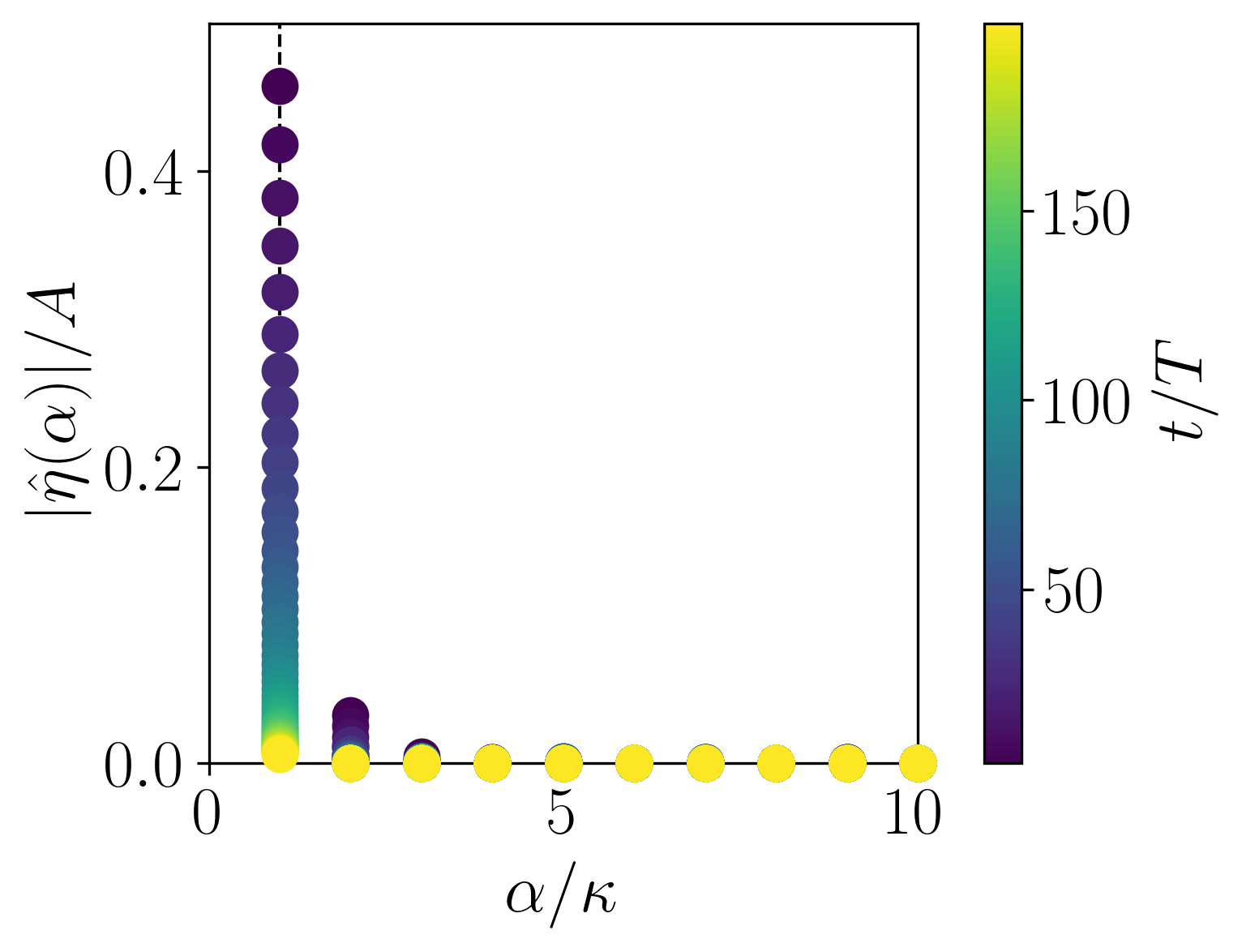}}
  \caption{{\textbf{(left)} Decay of the total energy $E$ over time from one of our simulations, compared to Landau's prediction. The inset, reproduced from \cite{foggirota-chiarini-rosti-2025}, reports the trend of a further validation case matching available literature \cite{iafrati-2009,devita-etal-2021-2}. {\textbf{(centre)}} Instantaneous and local dissipation rate $\Psi$ after $10T$ of our simulation with $Re=10000$ and $\epsilon=0.1$. Colours range linearly from null (black) to a maximum (white) of $\varepsilon \approx 100 U^3/\lambda $, and the interface position is denoted by a red curve. {\textbf{(right)} Un-normalised spectral components of the wave from the same simulation, throughout its decay.}}}
  \label{fig:decay}
\end{figure}

{
\subsection{Energy decay}
Given the sharp contrast between the simplicity of the assumptions in Landau's theory \cite{landau-lifshitz-2003} and the complexity of our simulated system, we assess its implications measuring the temporal decay of the total mechanical energy {(sum of the kinetic and potential energy of the water phase in our simulations)}.
The fall-off in the simulations is slightly faster than the theoretical prediction {$E(y)\sim e^{-2\gamma t}$, with $\gamma = 2 \nu_w \kappa^2$} (figure \ref{fig:decay}, left), {and we measure $\gamma_\mathrm{DNS} \approx 1.05 \gamma $}. Yet, the good agreement between our measurements with \textit{Fujin} and former simulation results \cite{iafrati-2009,devita-etal-2021-2} (validated in \cite{foggirota-chiarini-rosti-2025} and reported for completeness in the left inset of figure~\ref{fig:decay}) points towards the presence of further dissipative mechanisms compared to theory, rather than a fault of the numerics. The simulated energy decay remains nevertheless exponential (with good approximation), not affecting in principle the qualitative comparison of the theoretical predictions in the following with our measured trends.
}

{
To better understand the dissipative process, we compute the dissipation rate in the flow fields at different times (one of them is reported in the {centre} of figure~\ref{fig:decay}).
Dissipation within the fluid bulk always appears negligible, while most of it occurs in a thin air layer immediately above the surface. The reasonable agreement of Landau's decay prediction with our simulations is therefore impressive, noting that it completely overlooks the fact that the most dissipative process occurs in the air \cite{iafrati-babanin-onorato-2013,devita-etal-2021-2}.
}

{Finally, to elucidate the temporal evolution of the different spectral components excited by the wave, we report their un-normalised values for a selected case in the {right} of figure~\ref{fig:decay}.}

{
\subsection{Boundary effects}

\begin{figure}
\centering
  \hspace{1.8cm}
  \includegraphics[width=.22\textwidth]{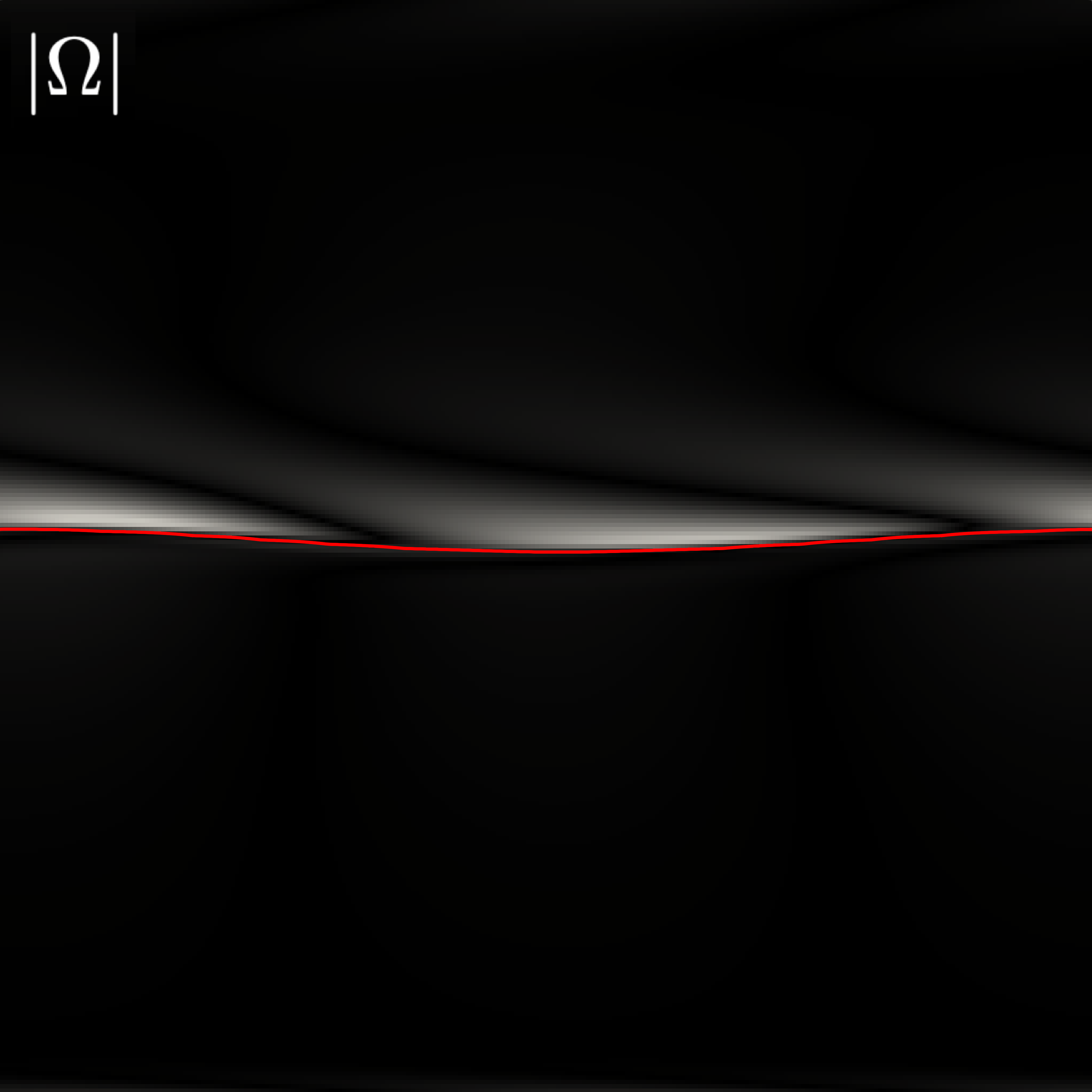}
\raisebox{-1cm}{%
   \includegraphics[width=.3\textwidth]{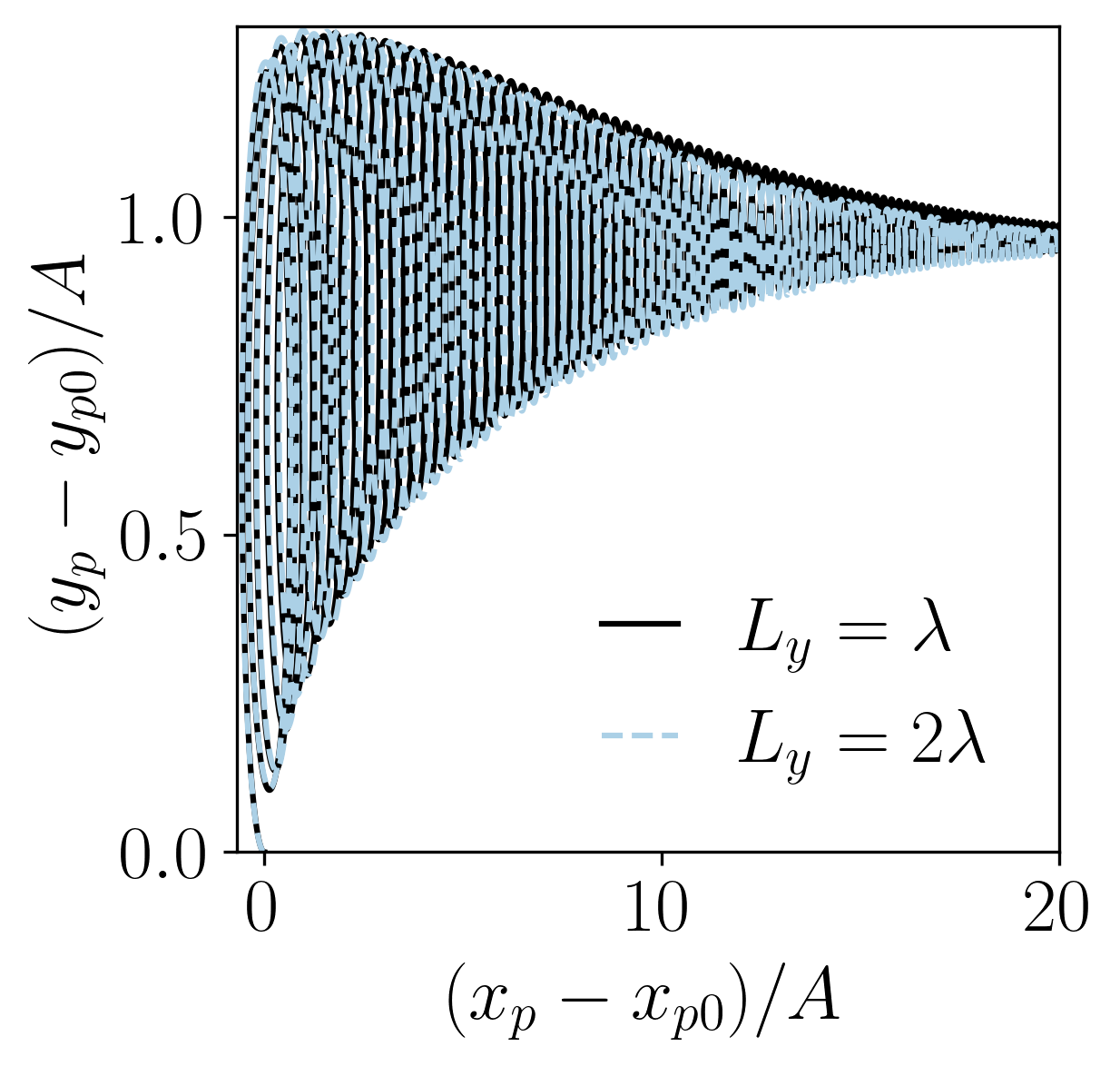}
}
\raisebox{-1cm}{%
  \includegraphics[width=.31\textwidth]{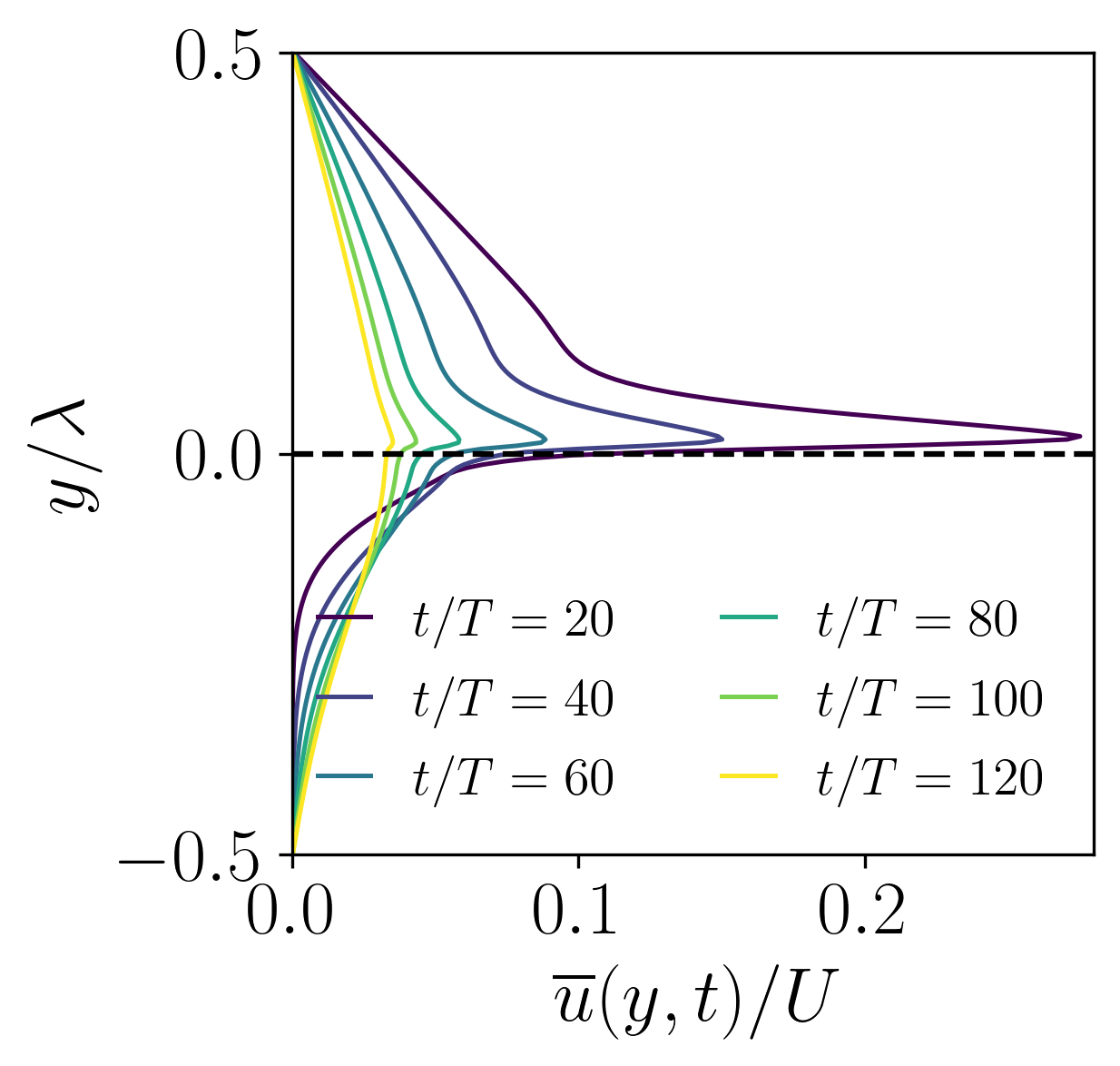}
}
  \caption{{\textbf{(left)} Vorticity magnitude $|\Omega|$ after $10T$ of our simulation with $Re=10000$ and $\epsilon=0.1$, ranging linearly from null (black) to a maximum (white) of $\varepsilon \approx 40 U/\lambda $. The interface position is denoted by a red curve. {\textbf{(centre)}} Trajectories of a particle released at $t=0$ with $\phi_p=0$ at a depth $y_p/A=-6$ with the same $Re$ and $\epsilon$, from the core of our study and from an additional case where the vertical extent of the domain has been doubled, keeping the interface in the middle. {\textbf{(right)} Averaged two-fluid velocity field $\overline{u}(y,t)$ at selected instants of the same simulation from the core of our study.}}}
  \label{fig:bl}
\end{figure}

To assess the effect of the top and bottom walls on the interface motion, we compute the vorticity in our flow fields at different times (one of them is reported in the left of figure~\ref{fig:bl}). Consistently with what previously observed for the dissipation, vorticity at the boundaries and within the bulk is always negligible, while most of it is concentrated over the surface. We thus notice that the water bulk remains essentially irrotational throughout the decay, consistently with the assumptions of our model. To further confirm the independence of our reported trends from viscous boundary effects, we replicate the simulation at $Re=10000$ and $\epsilon=0.1$ doubling the vertical extent of the domain, with the water surface in the middle. In the two cases, the trajectory of a selected particle released at $t=0$ with $\Phi_p=0$ at a depth $y_p/A=-6$ (where most of our measurements are performed) remains almost indistinguishable (in the {centre} of figure~\ref{fig:bl}). More appreciable discrepancies emerge close to the boundary.} 

{Since the numerical domain of our simulations is periodic along the direction of wave propagation, nothing prevents the establishment of a global mass flux in that direction.
Thus, differently from wave tank setups where boundaries induce recirculation, here we have a positive mass flux through the domain and no return at any depth.
Such effect is made evident by the averaged two-fluid velocity field $\overline{u}(y,t)$, reported for selected time instants of the case introduced above in the right of figure~\ref{fig:bl}.}

\section{The analytical model}

\subsection{Derivation}
\label{sec:derivation}

We begin by considering the inviscid limit to establish the basic wave dynamics. Specifically, we analyse gravity waves under the assumption that fluid velocity is small enough to neglect the nonlinear term $\bm{u} \cdot \bm{\nabla} \bm{u}$ in comparison to the unsteady term $\partial \bm{u} / \partial t$ in Euler’s equations. This corresponds to the condition:
\begin{equation}
A \ll \lambda,
\end{equation}
where $A$ is the wave amplitude and $\lambda$ is the wavelength. We take $x$ to be the streamwise direction and $y$ the vertical one, with the origin located at the undisturbed air–water interface.

In this regime, the velocity field is determined entirely by $A$, $\lambda$, and gravity $g$, and is given by
\begin{equation}
u(x,y,t) = U e^{\kappa y} \cos(\kappa x - \omega t), \qquad 
v(x,y,t) = U e^{\kappa y} \sin(\kappa x - \omega t),
\label{eq:pot-wave}
\end{equation}
where $\kappa = 2\pi/\lambda$ is the wavenumber, $U = \omega A$, and the dispersion relation $\omega^2 = \kappa g$ relates the wave frequency $\omega$ to $\kappa$. We also define the small parameter $\epsilon = \kappa A \ll 1$.

We now consider the effect of \textit{viscosity}. Following Landau \citep{landau-lifshitz-2003}, we assume the weakly viscous regime where
\begin{equation}
\nu_w \ll \lambda^2 \omega.
\end{equation}
Under this condition, vorticity is confined to a thin boundary layer near the surface, while the bulk flow remains approximately irrotational. Viscous dissipation leads to exponential decay of the wave amplitude $A e^{-\gamma t}$, where $\gamma = 2\nu_w \kappa^2$ is the damping rate. The mechanical energy of the wave thus decays as $e^{-2\gamma t}$ and the velocity field becomes:
\begin{equation}
u(x,y,t) = U e^{-\gamma t} e^{\kappa y} \cos(\kappa x - \omega t), \qquad
v(x,y,t) = U e^{-\gamma t} e^{\kappa y} \sin(\kappa x - \omega t).
\end{equation}

We now introduce dimensionless variables. For clarity, here dimensional quantities are denoted with a tilde (e.g., $\tilde{x}, \tilde{t}, \tilde{u}$), while un-tilded variables ($x, t, u$) are understood to be dimensionless:
\begin{equation}
\bm{u} = \frac{\tilde{\bm{u}}}{U}, \quad \bm{x} = \kappa \tilde{\bm{x}}, \quad t = \omega \tilde{t}, \quad \gamma = \frac{\tilde{\gamma}}{\omega}.
\end{equation}
The velocity field becomes:
\begin{equation}
u(x,y,t) = e^{-\gamma t} e^{y} \cos(x - t), \qquad
v(x,y,t) = e^{-\gamma t} e^{y} \sin(x - t).
\label{eq:dec-wave}
\end{equation}

We now analyze the motion of fluid particles. In dimensionless form the governing equations read
\begin{equation}
u_p(t) = \frac{\text{d} x_p}{\text{d} t} = \epsilon u(\bm{x}_p, t), \qquad
v_p(t) = \frac{\text{d} y_p}{\text{d} t} = \epsilon v(\bm{x}_p, t),
\label{eq:part}
\end{equation}
with initial condition $\bm{x}_p(0) = \bm{x}_{p_0}$.

We expand the particle position in powers of $\epsilon$:
\begin{equation}
x_p = x_0 + \epsilon x_1 + \epsilon^2 x_2 + \mathcal{O}(\epsilon^3), \qquad
y_p = y_0 + \epsilon y_1 + \epsilon^2 y_2 + \mathcal{O}(\epsilon^3),
\end{equation}
and use a Taylor expansion for $\bm{u}(\bm{x}_p,t)$, i.e.
\begin{equation}
\begin{gathered}
  u(\bm{x}_p,t) = u(\bm{x}_{p_0},t) + \left( x_p - x_{p_0} \right) \frac{\partial u}{\partial x}\big|_{\bm{x}_{p_0}} +
                                  \left( y_p - y_{p_0} \right) \frac{\partial u}{\partial y}\big|_{\bm{x}_{p_0}} + \mathcal{O}(|| \bm{x}_p - \bm{x}_{p_0} ||^2) \\
  v(\bm{x}_p,t) = v(\bm{x}_{p_0},t) + \left( x_p - x_{p_0} \right) \frac{\partial v}{\partial x}\big|_{\bm{x}_{p_0}} +
                                  \left( y_p - y_{p_0} \right) \frac{\partial v}{\partial y}\big|_{\bm{x}_{p_0}} + \mathcal{O}(|| \bm{x}_p - \bm{x}_{p_0} ||^2).                                 
\end{gathered}
\end{equation}
By injecting all this in the governing equation for the tracers we have:
\begin{equation}
  \begin{aligned}
  \dot{x}_0 + \epsilon \dot{x}_1 + \epsilon^2 \dot{x}_2 & = \epsilon u( \bm{x}_{p_0},t) + \epsilon \left( x_0 - x_{p_0} \right) \frac{ \partial u}{\partial x} \big|_{\bm{x}_{p_0}} +
                                                                                      \epsilon \left( y_0 - y_{p_0} \right) \frac{ \partial u}{\partial y} \big|_{\bm{x}_{p_0}} +
                                                                                      \epsilon^2 x_1 \frac{ \partial u}{\partial x} \big|_{\bm{x}_{p_0}} +
                                                                                      \epsilon^2 y_1 \frac{ \partial u}{\partial y} \big|_{\bm{x}_{p_0}} + \text{h.o.t.} \\
  \dot{y}_0 + \epsilon \dot{y}_1 + \epsilon^2 \dot{y}_2 & = \epsilon v( \bm{x}_{p_0},t) + \epsilon \left( x_0 - x_{p_0} \right) \frac{ \partial v}{\partial x} \big|_{\bm{x}_{p_0}} +
                                                                                      \epsilon \left( y_0 - y_{p_0} \right) \frac{ \partial v}{\partial y} \big|_{\bm{x}_{p_0}} +
                                                                                      \epsilon^2 x_1 \frac{ \partial v}{\partial x} \big|_{\bm{x}_{p_0}} +
                                                                                      \epsilon^2 y_1 \frac{ \partial v}{\partial y} \big|_{\bm{x}_{p_0}} + \text{h.o.t.}
  \end{aligned}
\end{equation}
where h.o.t. stands for higher order terms and dotted variables are differentiated in time. This leads to a hierarchy of equations that can be solved order by order. Here we truncate the expansion at $\mathcal{O}(\epsilon^2)$.

\begin{itemize}
\item At order $\epsilon^0$:
\begin{equation}
\dot{x}_0 = 0, \qquad \dot{y}_0 = 0 \quad \Rightarrow \quad x_0 = x_{p_0}, \quad y_0 = y_{p_0}.
\end{equation}
\item At order $\epsilon^1$:
\begin{equation}
\dot{x}_1 = u(\bm{x}_{p_0}, t), \qquad \dot{y}_1 = v(\bm{x}_{p_0}, t),
\end{equation}
with $x_1(0) = y_1(0) = 0$. Integration yields:
\begin{align}
x_1(t) &= \frac{e^{y_{p_0}}}{\gamma^2 + 1} \left[ -e^{-\gamma t} \sin(x_{p_0} - t) - \gamma e^{-\gamma t} \cos(x_{p_0} - t) \right] +
          {\frac{e^{y_{p_0}}}{\gamma^2 + 1} \left[                \sin(x_{p_0}    ) + \gamma               \cos(x_{p_0}    ) \right]} , \\
y_1(t) &= \frac{e^{y_{p_0}}}{\gamma^2 + 1} \left[ \phantom{-}e^{-\gamma t} \cos(x_{p_0} - t) - \gamma e^{-\gamma t} \sin(x_{p_0} - t) \right] +
          {\frac{e^{y_{p_0}}}{\gamma^2 + 1} \left[          -               \cos(x_{p_0}    ) + \gamma               \sin(x_{p_0}    ) \right]}
\end{align}
\item At order $\epsilon^2$:
\begin{align}
\dot{x}_2 &= x_1 \frac{\partial u}{\partial x} \bigg|_{\bm{x}_{p_0}} + y_1 \frac{\partial u}{\partial y} \bigg|_{\bm{x}_{p_0}} =  
            \frac{e^{-2\gamma t}}{\gamma^2 + 1} e^{2y_{p_0}} + 
            {\frac{e^{- \gamma t}}{\gamma^2 + 1} e^{2y_{p_0}} \left[ \gamma \sin(t) - \cos(t) \right]}, \\
\dot{y}_2 &= x_1 \frac{\partial v}{\partial x} \bigg|_{\bm{x}_{p_0}} + y_1 \frac{\partial v}{\partial y} \bigg|_{\bm{x}_{p_0}} = 
           -\frac{e^{-2\gamma t} \gamma}{\gamma^2 + 1} e^{2y_{p_0}} + 
            {\frac{e^{- \gamma t}}{\gamma^2 + 1} e^{2y_{p_0}} \left[ \gamma \cos(t) + \sin(t) \right]}.
\end{align}
\end{itemize}

Collecting all terms, the Lagrangian velocity becomes:
\begin{align}
u_p(t) &= \epsilon^1 \left[ e^{-\gamma t} e^{y_{p_0}} \cos(x_{p_0} - t) \right] +
          \epsilon^2 \left[ \frac{e^{-2\gamma t}}{\gamma^2 + 1} e^{2y_{p_0}} +
                            { \frac{ e^{-\gamma t} }{ \gamma^2 +1 } e^{2 y_{p_0}} \left( \gamma \sin(t) - \cos(t) \right) } \right] 
                            + \mathcal{O}(\epsilon^3) , \\
v_p(t) &= \epsilon^1 \left[ e^{-\gamma t} e^{y_{p_0}} \sin(x_{p_0} - t) \right] +
          \epsilon^2 \left[ - \frac{e^{-2\gamma t} \gamma}{\gamma^2 + 1} e^{2y_{p_0}} +
                            { \frac{ e^{-\gamma t} }{ \gamma^2 + 1} e^{2 y_{p_0}} \left( \gamma \cos(t) + \sin(t) \right) } \right]
                            + \mathcal{O}(\epsilon^3).
\label{eq:lagrangian-velocity-dimless}
\end{align}
We now return to dimensional variables (omitting tildes for brevity), and obtain:
\begin{align}
u_p(t) &= U e^{-\gamma t} e^{\kappa y_{p_0}} \cos(\kappa x_{p_0} - \omega t)
+ {\frac{U^2 \kappa}{\omega} \frac{e^{2\kappa y_{p_0}}}{(\gamma/\omega)^2 + 1} 
  \left[ e^{-2\gamma t} + e^{-\gamma t} \left( \frac{\gamma}{\omega} \sin(\omega t) - \cos(\omega t) \right) \right] } + \mathcal{O}(\epsilon^3).\\
v_p(t) &= U e^{-\gamma t} e^{\kappa y_{p_0}} \sin(\kappa x_{p_0} - \omega t)
+ {\frac{U^2 \kappa}{\omega} \frac{e^{2\kappa y_{p_0}}}{(\gamma/\omega)^2 + 1}
  \left[ -\frac{\gamma}{\omega} e^{-2 \gamma t} + e^{-\gamma t} \left( \frac{\gamma}{\omega} \cos(\omega t) + \sin(\omega t) \right) \right] } + \mathcal{O}(\epsilon^3).
\label{eq:velp}
\end{align}
This expression explicitly shows how viscous damping modifies the Lagrangian transport induced by surface gravity waves.

After integrating in time between $t_0$ and $t$ we obtain the $(x_p(t),y_p(t))$ trajectories of the tracers, being
\begin{equation}
  \begin{gathered}
    x(t) = x(t_0) + \frac{U}{\gamma^2 + \omega^2} e^{\kappa y_{p_0}} \Big[ 
    - e^{- \gamma t  } \left( \gamma \cos( \kappa x_{p_0} - \omega t  ) + \omega \sin( \kappa x_{p_0} - \omega t  ) \right) 
    + e^{- \gamma t_0} \left( \gamma \cos( \kappa x_{p_0} - \omega t_0) + \omega \sin( \kappa x_{p_0} - \omega t_0) \right) \Big] + \\
    + \frac{U^2 \kappa}{\gamma^2 + \omega^2} e^{2 \kappa y_{p_0}} \left[ - \frac{\omega}{2\gamma} \left( e^{-2 \gamma t} - e^{-2 \gamma t_0} \right) + { \left( - e^{-\gamma t} \sin(\omega t) + e^{- \gamma t_0 } \sin(\omega t_0) \right) } \right].
  \end{gathered}
  \label{eq:xtraj}
\end{equation}
and
\begin{equation}
  \begin{gathered}
    y(t) = y(t_0) + \frac{U}{\gamma^2 + \omega^2} e^{\kappa y_{p_0}} \Big[
      e^{- \gamma t  } \left( \omega \cos(\kappa x_{p_0} - \omega t  ) - \gamma \sin(\kappa x_{p_0} - \omega t  ) \right)
     -e^{- \gamma t_0} \left( \omega \cos(\kappa x_{p_0} - \omega t_0) - \gamma \sin(\kappa x_{p_0} - \omega t_0) \right) \Big] + \\
     + \frac{U^2 \kappa}{\gamma^2 + \omega^2} e^{2 \kappa y_{p_0}} \left[ \frac{1}{2} \left( e^{-2 \gamma t} - e^{-2 \gamma t_0} \right) + {\left( -e^{- \gamma t} \cos( \omega t ) + e^{- \gamma t_0} \cos(\omega t_0)  \right) } \right].
  \end{gathered}
  \label{eq:ztraj}
\end{equation}
Starting from equations \eqref{eq:xtraj} and \eqref{eq:ztraj}, we estimate the time evolution of the drift velocity. In particular, we define the drift velocity of a particle with initial position $\bm{x}_{p_0}$ at the $n$-th {wave} period as
\begin{equation}
    \bm{u}_d(\bm{x}_{p_0}, n) = \frac{\bm{x}_p(t_n) - \bm{x}_p(t_{n-1})}{T},
    \label{eq:veldrift}
\end{equation}
where $t_n = t_0 + nT$ and $n \in \mathbb{N}^+$. 

We conclude this section by examining the dependence of the asymptotic tracer displacement on the Reynolds number. As a first step, we set $t = t_0 + nT$ with $n \in \mathbb{N}^+$, and then take the limit of equations \eqref{eq:xtraj} and \eqref{eq:ztraj} as $n \rightarrow \infty$. This corresponds to taking the long-time limit $t \rightarrow \infty$, effectively averaging out the oscillatory phase. As a result, we obtain:
\begin{align}
  \lim_{t \rightarrow \infty} (x_p(t)-x_{p_0}) & = \frac{U e^{\kappa y_{p_0}} }{ \gamma^2 + \omega^2 } e^{-\gamma t_0} 
  \Big[ \gamma \cos(\kappa x_{p_0} - \omega t_0) + \omega \sin(\kappa x_{p_0} - \omega t_0) \Big] + \\
 & \frac{U^2 \kappa e^{2 \kappa y_{p_0}}}{\gamma^2 + \omega^2} e^{-\gamma t_0} \left[ \frac{\omega }{2\gamma} e^{-\gamma t_0} +
 { \sin(\omega t_0) } \right]   
\end{align}
and
\begin{align}
  \lim_{t \rightarrow \infty} (y_p(t)-y_{p_0}) & = \frac{U e^{\kappa y_{p_0}}}{\gamma^2 + \omega^2} e^{-\gamma t_0}
  \Big[ \gamma \sin(\kappa x_{p_0} - \omega t_0) - \omega \cos(\kappa x_{p_0} - \omega t_0) \Big] + \\
  & + \frac{U^2 \kappa e^{2 \kappa y_{p_0} } }{\gamma^2 + \omega^2} e^{-\gamma t_0} \left[ -\frac{ 1 }{2} e^{- \gamma t_0} + 
  { \cos(\omega t_0) } \right] 
\end{align}
At this point, we take the limit $\nu_w \rightarrow 0$ (i.e., $Re \rightarrow \infty$) to investigate the asymptotic dependence of $(x_p(t)-x_{p_0})$ and $(y_p(t)-y_{p_0})$ on the Reynolds number. We obtain:
\begin{equation}
  \lim_{t \rightarrow \infty,\nu \rightarrow 0} (x_p(t)-x_{p_0}) =   
  \frac{U e^{\kappa y_{p_0}} }{ \omega }  \sin(\kappa x_{p_0} - \omega t_0)
  + \frac{ U^2 \kappa e^{2 \kappa y_{p_0} } }{\omega}
  \left[ \underbrace{ \frac{1}{2 \gamma} }_{\sim Re^1} + { \frac{\sin(\omega t_0)}{\omega} } \right] \sim Re^1
\end{equation}
and
\begin{equation}
  \lim_{t \rightarrow \infty,\nu \rightarrow 0} (y_p(t)-y_{p_0}) = 
  - \frac{U e^{\kappa y_{p_0}}}{ \omega} \cos(\kappa x_{p_0} - \omega t_0)   
  + \frac{U^2 \kappa e^{2 \kappa y_{p_0}}}{\omega^2} \left[ - \frac{1}{2} + { \cos(\omega t_0) } \right] \sim Re^0.
\end{equation}

\subsection{Numerical integration}

\begin{figure}
  \centering
  \includegraphics[width=\textwidth]{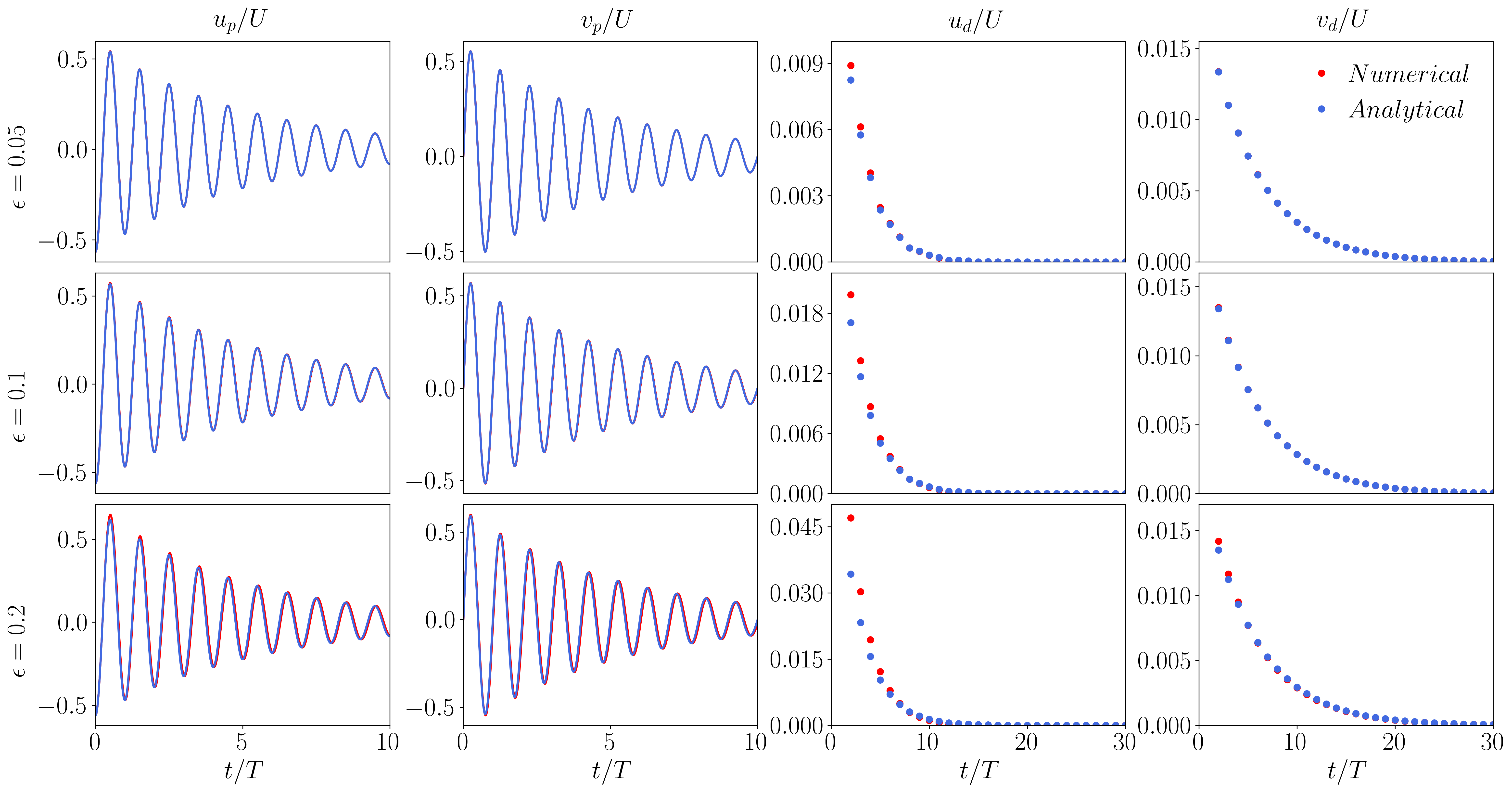}
  \caption{Comparison between model predictions (blue) and direct numerical integration (red) of equations \eqref{eq:part}, for the instantaneous tracer velocity $\bm{u}_p$ (equations \ref{eq:velp}) and drift velocity $\bm{u}_d$ (equation \ref{eq:veldrift}), for three values of the expansion parameter $\epsilon$. From left to right, the panels are for $u_p$, $v_p$, $u_d$ and $v_d$. From top to bottom the panels are for $\epsilon=0.05,0.1,0.2$. The red color refers to numerical integration, while the blue color to our analytical prediction.}
  \label{fig:traj-comp}
\end{figure}   
\begin{figure}
  \centering
  \includegraphics[width=\textwidth]{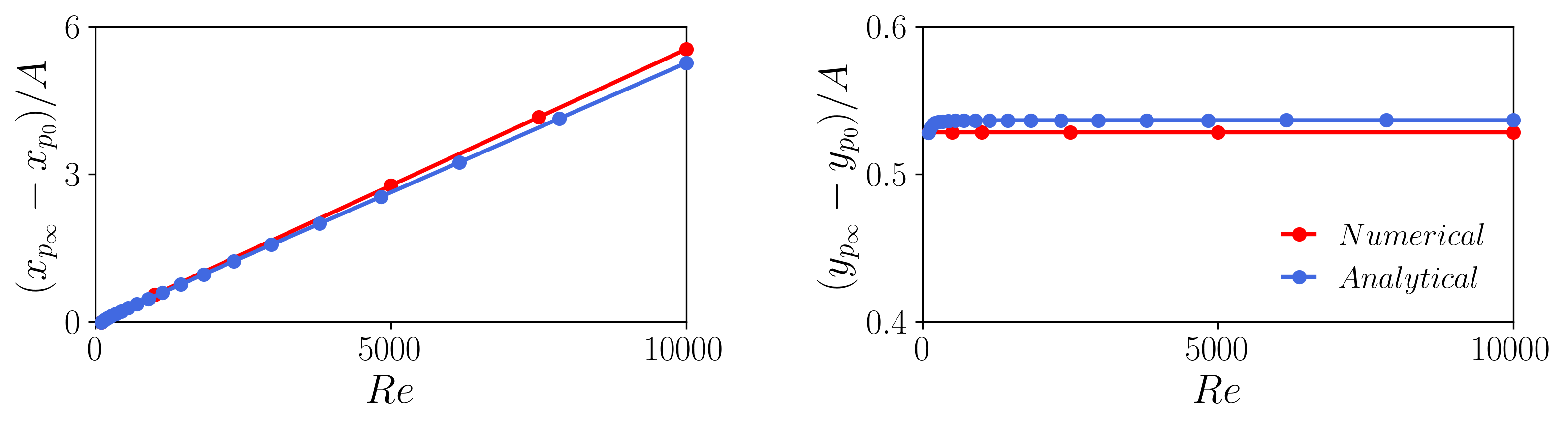}
  \caption{Asymptotic tracer displacement as a function of the Reynolds number. Longitudinal \textbf{(left)} and vertical \textbf{(right)} drift for a fluid element initially located at $\kappa x_{p_0} = \pi/2$ and $y_{p_0} = A \cos(\kappa x_{p_0} - \omega t_0) - 0.1$. Red symbols denote numerical integration, while blue symbols represent analytical predictions. The agreement remains good even for other initial positions. Here, $\epsilon = 0.0471$.}
  \label{fig:asymp}
\end{figure} 

In this section, we present results obtained integrating numerically the dimensional version of equations \eqref{eq:part}, with the goal of verifying the analytical predictions derived above and assessing their robustness with respect to the expansion parameter~$\epsilon$.

We fix the parameters to $\lambda = 1$, $g = 1$, $\nu_w = 1/1000$, $\omega ={ \sqrt{2\pi}}$, and vary the initial wave amplitude~$A$ to explore different values of~$\epsilon$. To evaluate the accuracy of our predictions, we begin analysing the time-dependent tracer velocity $\bm{u}_p$ and the period-dependent drift velocity $\bm{u}_d$, see equations \eqref{eq:velp} and \eqref{eq:veldrift}. {The numerical estimate of $\bm{u}_d$ is computed by substituting the integrated tracer positions at times $t_n$ and $t_{n-1}$ into equation \eqref{eq:veldrift}}.

Figure \ref{fig:traj-comp} shows that the analytical predictions agree well with the numerical results when the expansion parameter $\epsilon$ remains small. As expected, for $\epsilon > 0.1$, the agreement deteriorates due to the breakdown of the perturbative approximation. However, as the wave amplitude decays over time and the instantaneous value of $\epsilon$ decreases, the agreement improves significantly. 


%
Furthermore, in figure \ref{fig:asymp}, we compare the asymptotic tracer displacement as a function of the Reynolds number. Once again, the numerical integration confirms the analytical predictions. In the limits $t \rightarrow \infty$ and $Re \rightarrow \infty$, the longitudinal drift $(x_{p_\infty}-x_{p_0})$ exhibits a linear dependence on $Re$, while the vertical drift $(y_{p_\infty}-y_{p_0})$ approaches an asymptotic value that depends on the initial conditions rather than on $Re$.


\subsection{Comparison with the simulations}

\begin{figure}[b]
   \centering
  \includegraphics[width=\textwidth]{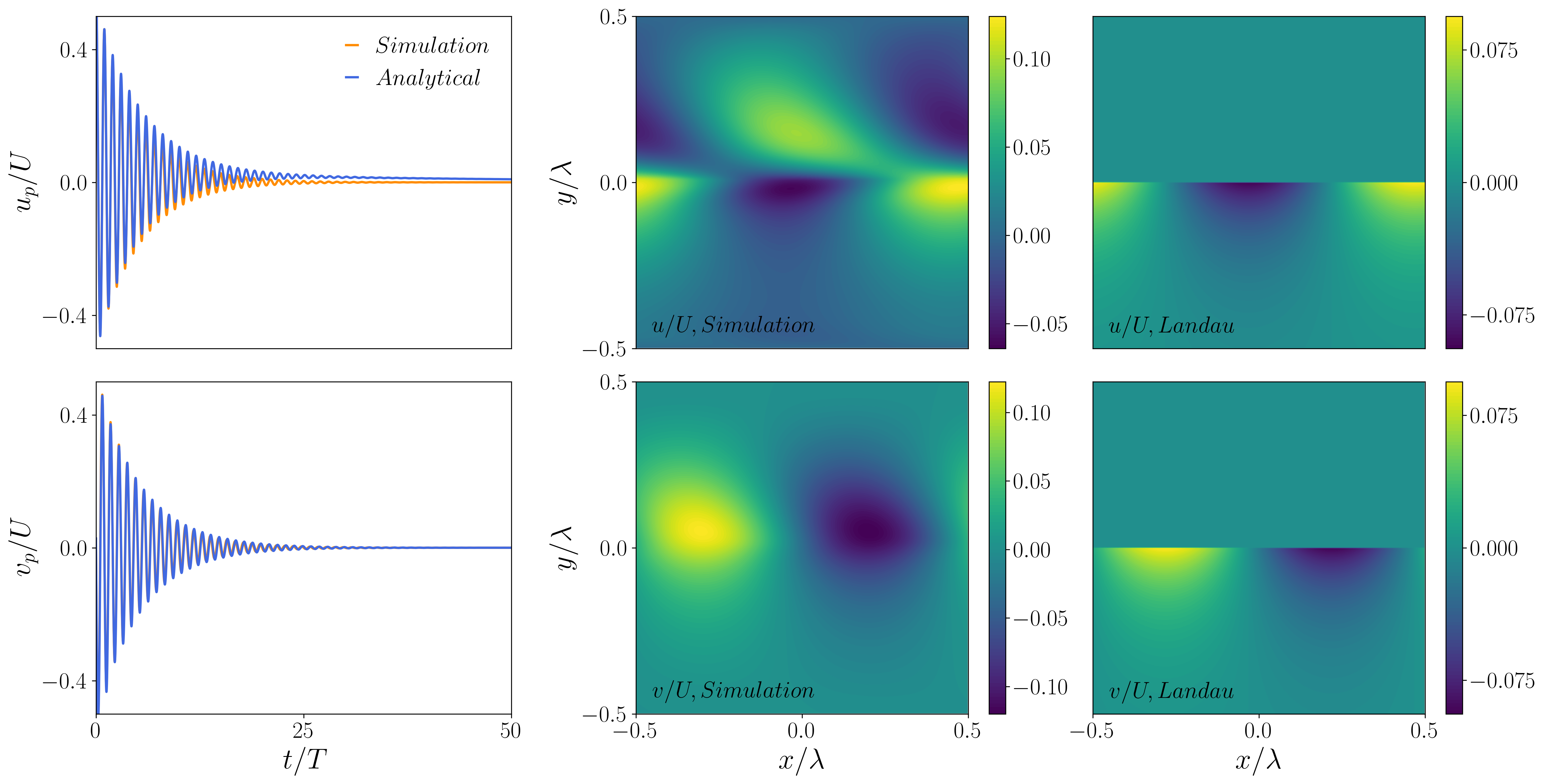}
  \caption{\textbf{(left)} Comparison between analytical predictions (blue) and simulations (orange) of the instantaneous velocity components of $\mathbf{u_p}{=\{u_p,v_p\}}$ {(top and bottom, respectively)} of a tracer initially positioned at $x_{p_0} / \lambda = 0$ and $y_{p_0} / \lambda= -0.09$. We consider $\nu = 1/1000$, $\lambda = 1$, $g = 1$, and $\epsilon = 0.1$. \textbf{(right, 4 panels)} Comparison of the flow field {components ($u$/$v$ in the top/bottom, respectively)} predicted by Landau's theory and computed with our simulation for $\epsilon=0.1$ at $t/T=12$, {as in captions}. In both cases, at $t=0$ the flow field is initialised with the potential solution \eqref{eq:pot-wave}. While at small times the flow fields are similar, already at moderately larger times {-- like that chosen for this plot --} the difference between the fields is evident.}
  \label{fig:dns-mod-comp}
\end{figure}

%
%

In this subsection, we compare the model predictions with the fully viscous and nonlinear direct numerical simulations. In the previous subsection, we showed that our perturbative expansion agrees well with the numerical integration of equations \eqref{eq:part} based on the weakly viscous assumptions of Landau, provided that the expansion parameter remains small $\epsilon \lessapprox 0.1$. Here, by directly comparing the model predictions with our simulations, we assess the validity of Landau's assumptions regarding the decay of the local velocity field. While the decay of wave energy is well captured by Landau's theory~\citep{iafrati-2009,devita-etal-2021}, its accuracy in describing the local flow field---and, by extension, Lagrangian transport---has not been assessed yet.


The {two} left panels of figure \ref{fig:dns-mod-comp} compare the instantaneous velocity {components} of a tracer, initially positioned at $x_{p_0}/\lambda = 0$ and $y_{p_0} / \lambda = -0.09$, as predicted by the model and computed via simulations. We consider $\nu_w = 1/1000$, $\lambda = 1$, $g = 1$, and $\epsilon = 0.1$. At early times, the agreement is good: for both $u_p$ and $v_p$, the signals overlap closely, with the theoretical model slightly under-predicting the peak amplitudes.
%
At later times, however, discrepancies emerge. While $v_p$ continues to be well captured, the model prediction for $u_p$ deviates from the DNS results, resulting in a systematic underestimation of the longitudinal drift velocity at large times at several $Re$ (not shown). Since the agreement is strong during the initial oscillation periods---with the drift velocity closely matching between simulations and model---we attribute the growing deviation to a progressive divergence between the simulated flow field and that assumed in the Landau-based model, {shown in the four right panels of} figure~\ref{fig:dns-mod-comp}. Even small differences between the predicted and simulated tracer velocities induce significant divergence of their trajectories over long times.
{The divergence can instead not be inputed to the slight difference between the theoretical decay exponent $\gamma$ and the measured trend $\gamma_\mathrm{DNS}$, as it persists even when properly compensating to match the two.}

In conclusion, while the theoretical model and the simulations remain in qualitative agreement, quantitative differences become apparent at long times, being particularly evident in the asymptotic behaviour for $t \rightarrow \infty$. These discrepancies in the flow field likely account for the model's reduced accuracy in predicting long-time Lagrangian transport.



%